\documentstyle[pre,aps,preprint]{revtex} 
\begin{document}                                     
%\draft
\title{Gas-kinetic derivation of Navier-Stokes-like traffic equations}
\author{Dirk Helbing}
\address{II. Institute of Theoretical Physics, University of
Stuttgart, 70550 Stuttgart, Germany}
\maketitle                                                    
\begin{abstract}
Macroscopic traffic models have recently been severely criticized to base on
lax analogies only and to have a number of deficiencies. Therefore, this paper
shows how to construct a logically consistent fluid-dynamic traffic model
from basic laws for the acceleration and interaction of vehicles. These
considerations lead to the gas-kinetic traffic equation of Paveri-Fontana.
Its stationary and spatially homogeneous solution implies equilibrium
relations for the `fundamental diagram', the variance-density 
relation, and other quantities which are partly difficult to determine 
empirically.
\par
Paveri-Fontana's traffic equation allows the derivation of macroscopic
moment equations which build a system of non-closed equations. This system
can be closed by the well proved method of Chapman and Enskog which leads
to Euler-like traffic equations in zeroth-order approximation and to
Navier-Stokes-like traffic equations in first-order approximation. The latter
are finally corrected for the finite space requirements
of vehicles. It is shown that the resulting model is able to
withstand the above mentioned criticism.
\end{abstract}                                                
%\begin{multicols}{2}                           
\section{Introduction}

Because of analogies with gas theory \cite{Nels,Pav,Phil,Prig} and 
fluid dynamics \cite{Hel1,Hill,HillWeid,Kern1,Kern2,Phil,LW} 
modeling and simulating traffic
flow increasingly attracts the attention of physicists 
\cite{Nels,Hel1,Kern1,Kern2,Nagel,PRE1,PRE2,PRE3}.
However, due to the great importance of efficient traffic for modern
industrialized countries, the investigation of traffic flow has already a long
tradition. In the 1950s Lighthill and Whitham \cite{LW} as well as
Richards \cite{Rich} proposed a first {\em fluid-dynamic (macroscopic)
traffic model}. During the 1960s
traffic research focused on {\em microscopic follow-the-leader models}
\cite{FL5,FL6,FL7,FL8,FL1,FL2,FL3,FL4}. {\em Mesoscopic models} of a 
{\em gas-kinetic (Boltzmann-like) type}
came up in the 1970s \cite{Andr,Pri,Prig,Phil,Pav,Alb}. Since the 1980s 
{\em simulation models} \cite{SIM1,SIM2} play
the most important role due to the availability of cheap, fast, and powerful
computers. We can distinguish {\em macroscopic traffic simulation models}
\cite{FRE,Cre1,Kuehne1,Kern3}, {\em microscopic simulation models}
\cite{MIC1,Netsim,MIC2,MIC3} 
which include {\em cellular automaton models}
\cite{Cre2,Nag1,Nag2,Nagel,PRE1,PRE2,PRE3}, and mixtures of both 
\cite{Dynemo}. 
\par
In {\em high-fidelity} microscopic traffic models each car is described by its
own equation(s) of motion. Consequently, computer time and memory requirements
of corresponding traffic simulations grow proportional to the number $N$ of
simulated cars. Therefore, this kind of models is mainly suitable for off-line
traffic simulations, detail studies (for example of on-ramps or lane
mergings), or the numerical evaluation of collective quantities \cite{MIC1}
like the
density-dependent velocity distribution, the distribution of headway distances
etc., and other quantities that are difficult to determine empirically.
\par
For this reason, fast {\em low-fidelity} microsimulation models that allow
bit-handling have been developed for the simulation of large freeways
or freeway networks \cite{Cre2,Nag1}. However, although they 
reproduce the main effects of traffic flow, they are not
very suitable for detailed {\em predictions} because of
their coarse-grained description. 
\par
Therefore, some authors prefer {\em macroscopic traffic models}
\cite{LW,Payne,Papa,Kuehne,Cre1,Smu,Hel1,Hill,HillWeid,Kern1,Kern2}. 
These base on equations for collective quantities
like the average spatial {\em density} $\rho(r,t)$ per lane
(at place $r$ and time $t$),
the {\em average velocity} $V(r,t)$, and maybe also the velocity {\em variance}
$\Theta(r,t)$. Here, simulation time and memory requirements mainly depend
on the discretization $\Delta r$ and $\Delta t$ of space $r$ and time $t$,
but not on the number $N$ of cars. Therefore, macroscopic traffic models
are suitable for {\em real-time} traffic simulations. The quality and
reliability of the simulation results mainly depend on the
{\em correctness} of the applied macroscopic equations and the choice of a
suitable {\em numerical integration method}. The rather old and still
continuing controversy on these problems 
\cite{Payne,Payne2,Hauer,Papa,Kuehne,Bab,Cre1,Smu,Lieber,Ross,New,Kern1,Kern2,Hel1,Nels,Daganzo} 
shows that they are not at all trivial. 
\par
Some of the most important points of this controversy will be %shortly
outlined in Section II. It will be shown that even the most advanced models
still have some serious shortcomings. The main reason for this is that the
proposed macroscopic traffic equations were founded on heuristic arguments
or based on analogies with the equations for ordinary fluids.
In contrast to these approaches, this paper will present a {\em mathematical 
derivation} of macroscopic traffic equations starting from the
gas-kinetic {\em traffic equation of Paveri-Fontana} \cite{Pav} which is very
reasonable and seems to be superior to the one of Prigogine and co-workers
\cite{Andr,Pri,Prig}. The applied method
is analogous to the derivation of the {\em Navier-Stokes equations} for
ordinary fluids from the {\em Boltzmann equation} 
\cite{Jaeckle,Rieck,Huang,Lib}. It bases on 
a {\em Chapman-Enskog expansion} \cite{Chap,Ensk} 
which is known from kinetic gas theory and leads 
to idealized, {\em Euler-like equations} in zeroth-order approximation 
and to {\em Navier-Stokes-like equations} in 
first-order approximation \cite{Rem2,Lib}. In this
respect, the paper puts into effect the method suggested 
by Nelson \cite{Nels}. A similar method was already applied to the derivation
of fluid-dynamic equations for the motion of pedestrian
crowds \cite{Hel2}, but it assumed some dissatisfactory approximations.
\par
The further procedure of this paper is as follows: Section II presents a short
history of macroscopic traffic models and discusses the abilities and
weaknesses of the different approaches. Section III introduces the
Boltzmann-like model of Prigogine \cite{Prig} and compares it
with the one of 
Paveri-Fontana \cite{Pav}. From their gas-kinetic equations macroscopic 
(`fluid-dynamic') traffic equations will be derived in Section IV. 
Unfortunately, they turn out to build a {\em hierarchy of
non-closed equations}, i.e. the density equation depends on average
velocity $V$, the velocity equation on velocity variance $\Theta$, etc.
Therefore, a suitable approximation must be found to obtain a set of 
closed equations. It will be shown that some of the traffic models introduced
in section II correspond to {\em zeroth-order approximations} of different 
kinds. These, however, are not very well justified. 
A similar thing holds for the Euler-like traffic
equations which, apart from a complementary covariance equation, 
contain additional terms compared with the Euler equations of
ordinary fluids \cite{Rem2}. These are, on the one hand, 
due to a {\em relaxation term} which describes
the drivers' acceleration towards their desired velocities. On the
other hand, they are due to 
interactions which are connected with deceleration processes
since these do not satisfy momentum and energy conservation
in contrast to atomic collisions. 
\par
A very realistic, {\em first-order approximation} which
is, in a certain sense, self-consistent can be found by solving
the reduced Paveri-Fontana equation which is obtained from the original
one by integration with respect to desired velocity. 
We will utilize the fact that, according to
empirical traffic data \cite{Pampel,MuPi,Phil,Notiz,MIC1}, 
the equilibrium velocity distribution has a Gaussian form.
This allows the derivation of mathematical expressions for
the equilibrium velocity-density relation, the `fundamental diagram' of
traffic flow, and the equilibrium variance-density
relation (cf. Sec. IV C). 
Afterwards an approximate time-dependent solution of 
Paveri-Fontana's equation will be calculated by use of the
Euler-like equations. Due to the additional terms in
Paveri-Fontana's equation compared with the Boltzmann equation 
the corresponding mathematical procedure is more complicated 
than the Chapman-Enskog expansion for ordinary gases (cf. Sec. V). 
\par
Nevertheless, it is still possible to derive correction terms of the 
Euler-like macroscopic traffic equations (cf. Sec. VI). 
These have the meaning of {\em transport terms} (like e.g. the
flux density of velocity variance) and are
related with the finite {\em skewness} $\gamma$ of the velocity
distribution in non-equilibrium situations.
The resulting equations are Navier-Stokes-like traffic equations which,
in comparison with the ordinary Navier-Stokes equations \cite{Rem2},
contain additional terms arising from the acceleration and interaction 
of vehicles. Additionally, 
they are complemented by a covariance equation which takes
into account the tendency of drivers to adapt to their desired velocities.
\par
Because of the one-dimensionality of the Navier-Stokes-like
traffic equations no {\em shear viscosity} term occurs. 
However, in Section VII it is indicated how transitions
between different driving modes can cause a {\em bulk viscosity} term.
Furthermore, corrections due to finite space requirements of each
vehicle (vehicle length plus safe distance) are introduced.
\par
The resulting model overcomes the shortcomings of the former 
macroscopic traffic models (that are mentioned in Sec. II).
Section VIII summarizes the results of the paper and gives a short
outlook.

\section{Short history of macroscopic traffic models}

In 1955 Lighthill and Whitham \cite{LW}
proposed the first macroscopic (fluid-dynamic)
traffic model. This bases on the {\em continuity equation}
\begin{equation}
 \frac{\partial \rho}{\partial t} + \frac{\partial (\rho V)}{\partial r} = 0
\label{cont}
\end{equation}
which reflects a conservation of the number of vehicles. For the average
velocity $V$, Lighthill and Whitham assumed
a static velocity-density relation:
\begin{equation}
 V(r,t) := V_{e}[\rho(r,t)] \, .
\label{LW}
\end{equation} 
Inserting (\ref{LW}) into (\ref{cont}) we obtain
\begin{equation}
 \frac{\partial \rho}{\partial t} + \left[ V_{e} + \rho
 \frac{\partial V_{e}}{\partial \rho} \right] \frac{\partial \rho}{\partial
  r} = 0 \, .
\label{resu}
\end{equation}
Equation (\ref{resu}) describes the propagation of non-linear 
{\em `kinematic waves'} with velocity $c(\rho) = V_{e}(\rho)
+ \rho \, \partial V_{e} / \partial \rho$ \cite{LW,Whith}. 
%These waves are
%marginally stable, i.e. small deviations from the
%stationary and homogeneous solution are neither amplified nor damped.
In the course of time the waves develop a {\em shock structure}, i.e. their 
back becomes steeper and steeper until it becomes perpendicular, leading to
discontinuous wave profiles \cite{LW,Rich,Whith}. 
\par
In reality, density changes are not so extreme. Therefore, it was suggested to
add a {\em diffusion term} $D \partial^2 \rho/\partial r^2$ which smoothes 
out the shock structures somewhat \cite{Whith,Musha}. 
The resulting equation reads
\begin{equation}
 \frac{\partial \rho}{\partial t} + V_{e} \frac{\partial \rho}{\partial r}
 = - \rho \frac{\partial V_{e}}{\partial \rho} \frac{\partial \rho}
 {\partial r} + D \frac{\partial^2 \rho}{\partial r^2} \, .
\label{Burg}
\end{equation}
For the case of a linear velocity-density relation \cite{Green}
\begin{equation}
 V_{e}(\rho) := V_{\rm max} \left( 1 - \frac{\rho}{\rho_{\rm max}} \right)
\end{equation}
it can be transformed into the {\em Burgers equation} \cite{Burger}
\begin{equation}
 \frac{\partial g}{\partial t} + g \frac{\partial g}{\partial r}
 = D \frac{\partial^2 g}{\partial r^2} 
\end{equation}
which is analytically solvable \cite{Whith}. Here, we have introduced the
function
\begin{equation}
 g[\rho(r,t)] := V_{\rm max} \left( 1 - \frac{2\rho(r,t)}
 {\rho_{\rm max}} \right)  \, .
\end{equation}
\par
The most important restriction of models (\ref{cont}), (\ref{LW})
and (\ref{Burg}), (\ref{LW}) is relation (\ref{LW}) which assumes that 
average speed $V(r,t)$ is always in equilibrium with density $\rho(r,t)$.
Therefore, these models are not suitable for the description 
of non-equilibrium 
situations occuring at on-ramps, changes of the number of lanes, or
stop-and-go traffic. 
\par
Consequently, it was suggested to replace relation (\ref{LW}) by a dynamic
equation for the average velocity $V$. In 1971, Payne \cite{Payne} introduced 
the {\em velocity equation}
\begin{mathletters}\label{Payne}
\begin{equation}
 \frac{\partial V}{\partial t} + V\frac{\partial V}{\partial r}
 = - \frac{C(\rho)}{\rho} \frac{\partial \rho}{\partial r}
 + \frac{1}{\tau} [ V_{e}(\rho) - V ] 
\end{equation}
with
\begin{equation}
 C(\rho) := - \frac{1}{2 \tau} \frac{\partial V_{e}}{\partial \rho} 
 = \frac{1}{2 \tau} \left| \frac{\partial V_{e}}{\partial \rho} \right| 
\label{Ce}
\end{equation}\end{mathletters}
which he motivated by a heuristic derivation from a microscopic
follow-the-leader model \cite{Note1}.
Here, $V \partial V/\partial r$ is called the {\em `convection term'} and
describes velocity changes at place $r$ that are caused by average
vehicle motion. The {\em `anticipation term'}
$- (C/\rho) \partial \rho/\partial r$ was intended to account for 
the drivers' awareness of the traffic conditions ahead. 
Finally, the {\em `relaxation term'} 
$[ V_{e}(\rho) - V ] / \tau$ delineates an (exponential) adaptation
of average velocity $V$ to the {\em equilibrium velocity} $V_{e}(\rho)$
with a {\em relaxation time} $\tau$.
\par
Unfortunately, for bottlenecks
the corresponding computer simulation program 'FREFLO' suggested
by Payne \cite{FRE} produces output that ``does not seem to reflect what
really happens even in a qualitative manner'' \cite{Hauer}.
%predicts results for bottlenecks that do not 
%not even qualitatively meaningful \cite{...}.
As a consequence, several authors have suggested a considerable number of
modifications of Payne's numerical integration method or of his equations
\cite{Note2,Papa,Kuehne,Bab,Cre1,Smu,Lieber,Ross,Leo}. A more principal weakness
of Payne's equations is that their stationary and homogeneous solution
is stable with respect to fluctuations  over the whole density range
which can be shown by a linear stability analysis \cite{Note2,Payne2,Payne}. 
Therefore, Payne's model (\ref{cont}), (\ref{Payne})
does not describe the well-known self-organization of
stop-and-go waves above a critical density \cite{Kuehne,Kuehne2}. This problem
is removed \cite{Payne2} by substituting relation (\ref{Ce}) by 
\begin{equation}
 C(\rho) := \frac{\partial {\cal P}_{e}}{\partial \rho}
\end{equation}
with the {\em equilibrium `traffic pressure'} 
\begin{equation}
 {\cal P}_{e}(\rho) := \rho \Theta_{e}(\rho) \, .
\end{equation}
The modified velocity equation reads
\begin{mathletters}\label{Phillips}
\begin{equation}
 \frac{\partial V}{\partial t} + V\frac{\partial V}{\partial r}
 = - \frac{1}{\rho} \frac{\partial {\cal P}_{e}}{\partial r}
 + \frac{1}{\tau} [ V_{e}(\rho) - V ]
\label{vel} 
\end{equation}
and can be derived from the gas-kinetic (Boltzmann-like) traffic models
\cite{Prig,Phil,Pav} (cf. Section IV). For $\Theta_{e}(\rho)$, 
Phillips \cite{Phil,Phil2}
suggested a relation of the form
\begin{equation}
 \Theta_{e}(\rho) :=
 \Theta_0 \left( 1 - \frac{\rho}{\rho_{\rm max}} \right) \, .
\label{Philb}
\end{equation}\end{mathletters}
In contrast, K\"uhne \cite{Kuehne3} as well as Kerner 
and Konh\"auser \cite{Kern1,Kern2} %, due to a
%lack of information about the functional form of $\Theta$, 
assumed, as a first approach, $\Theta_{e}$ to be a positive constant:
\begin{equation}
 \Theta_{e}(\rho) := \Theta_0 \, .
\end{equation}
Unfortunately, equations (\ref{cont}), (\ref{vel}) predict the formation
of shock waves like Lighthill and Whitham's equation does \cite{Kuehne,Hel1}. 
For this reason, K\"uhne \cite{Kuehne,Kuehne2}
suggested to add a small viscosity term $\nu \partial^2 V/\partial r^2$
which smoothes out sudden density and velocity changes somewhat.
Then, the velocity equation 
\begin{equation}                                                
 \frac{\partial V}{\partial t} + V\frac{\partial V}{\partial r}
 = - \frac{\Theta_0}{\rho} \frac{\partial \rho}{\partial r} 
 + \nu \frac{\partial^2 V}{\partial r^2} + \frac{1}{\tau} [ V_{e}(\rho) - V ]
\label{kuehne} 
\end{equation}
results. A linear stability analysis of 
K\"uhne's equations (\ref{cont}), (\ref{kuehne})
shows that these predict the self-organization of
{\em stop-and-go waves} or of so-called {\em `phantom traffic jams'}
(i.e. unstable traffic) on the condition
\begin{equation}
 \rho_e \left| \frac{\partial V_{e}}{\partial \rho} \right| >
 \sqrt{\Theta_0} (1 + \tau \nu k^2) 
\label{cr}
\end{equation}
where $k$ denotes the wave number of the perturbation \cite{Kuehne4,Hel1}.
This condition is fulfilled if the equilibrium density $\rho_e$ 
corresponding to the stationary and spatially homogeneous solution exceeds a 
{\em critical density} $\rho_{\rm cr}$ that depends on the concrete
form of $V_{e}(\rho)$. 
\par
For reasons of compatibility with the Navier-Stokes equations for ordinary
fluids Kerner and Konh\"auser replaced K\"uhne's constant $\nu$ by
the density-dependent relation
\begin{equation}
 \nu(\rho) = \frac{\nu_0}{\rho}
\label{eta}
\end{equation}
with the constant {\em viscosity coefficient} $\nu_0$. Computer simulations
of their equations (\ref{cont}) and (\ref{kuehne}), (\ref{eta})
show the development of {\em density clusters} 
\cite{Kern1,Kern2} if the critical
density $\rho_{\rm cr}$ given by (\ref{cr}) and (\ref{eta}) is 
exceeded. On the basis of a very comprehensive study of cluster-formation
phenomena, Kerner and Konh\"auser \cite{Kern2} presented a detailed 
interpretation of stop-and-go traffic.
\par
Despite the considerable variety of proposed macroscopic traffic models,
even the most advanced of them have still some shortcomings.
For example, for a certain set of parameters the mentioned models predict
traffic densities that exceed the maximum admissible density
$\rho_{\rm bb} = 1/l_0$ which is the {\em bumper-to-bumper density}
($l_0 = \mbox{average}$ vehicle length) \cite{Hel1}. 
Furthermore, in certain situations even
negative velocities may occur \cite{Daganzo}. To illustrate this, imagine a 
queue of vehicles of constant density $\rho_0$. Assume that,
e.g. due to an accident that blocks the road, this queue has come to rest  
(i.e. $V = 0$) and that it ends at $r = r_0$ which shall imply $\rho(r,t) = 0$ 
for $r < r_0$. Then, $\partial \rho/\partial r$ diverges 
at place $r_0$ (or is at least very
large) and equations (\ref{Payne}), (\ref{Phillips}), (\ref{kuehne}) 
all predict $\partial V(r_0,t)/\partial t < 0$ if $\Theta \ne 0$.
\par
Of course, we wish to have a model that is not only valid in standard
situations, but also in extreme ones. Moreover, the model should provide
reasonable results not only for certain parameter values. This is particularly
important for the reason that technical measures like automatic distance
control may change some parameter values considerably. Nobody knows if 
the existing phenomenological models are still applicable, then. Therefore, we
will derive the specific structure of the traffic  model from basic principles
regarding the behavior of the single driver-vehicle units and their
interactions. 

\section{Gas-kinetic (Boltzmann-like) traffic models}

Let us assume that the motion of an individual vehicle $\alpha$ can be
described by several variables like its place $r_\alpha
(t)$, its velocity
$v_\alpha(t)$, and maybe other quantities which characterize the vehicle type
or driving
style (the driver's personality). We can combine these quantities in a vector
\begin{equation}
 \vec{x}_\alpha(t) := \Big( r_\alpha(t),v_\alpha(t),\dots \Big)
\end{equation}
that denotes the {\em state} of vehicle $\alpha$ at a given time t.
The time-dependent {\em phase-space density}
\begin{equation}
 \hat{\rho}(\vec{x},t) \equiv \hat{\rho}(r,v,\dots,t)
\end{equation}
is then determined by the mean number $\Delta 
n(r,v,\dots,t')$ of vehicles that are at a
place between $r - \Delta r/2$ and $r + \Delta r/2$, 
driving with a velocity between $v - \Delta v/2$ and $v + \Delta v/2$, $\dots $
at a time $t' \in [t - \Delta t/2,t + \Delta t/2]$:
\begin{equation}
 \hat{\rho}(r,v,\dots,t)\, \Delta r \, \Delta v \dots 
 := \frac{1}{\Delta t} \int\limits_{t-\Delta t/2}^{t + \Delta t/2}
 \!\!\!\! dt' \, \Delta n(r,v,\dots,t') \, .
\end{equation}
For vehicles, the phase-space densitiy $\hat{\rho}$ is a very small 
quantity. Therefore, in the limit $\Delta r \rightarrow 0$,
$\Delta v \rightarrow 0$, $\dots$, $\Delta t \rightarrow 0$
it is only meaningful in the sense of the {\em expected value}
of an {\em ensemble} of macroscopically identical systems \cite{Nels}.
The interpretation of $\hat{\rho}$
as a quantity which can describe single traffic
situations is only possible for {\em ``coarse-grained averaging''}
where $\Delta r$, $\Delta v$, $\dots$, and $\Delta t$ must be chosen
``microscopically large but macroscopically small'' \cite{Nels,Hel2} or,
more exactly,
\begin{itemize}
\item[1.] smaller than the scale on which variations of the 
corresponding macroscopic quantities occur,
\item[2.] so large that $\Delta n(r,v,\dots,t) \gg 1$ which is not always
compatible with the first condition. 
\end{itemize}
However, in any case a suitable gas-kinetic equation for the phase-space density
$\hat{\rho}$ allows the derivation of meaningful equations for collective
(`macroscopic') quantities like the spatial density $\rho(r,t)$ per lane, the
average velocity $V(r,t)$, and the velocity variance $\Theta(r,t)$.
%{\em density} 
%\begin{equation}
% \rho(r,t) = \int \dots \int dv \, \hat{\rho}(r,v,\dots,t) \, ,
%\end{equation}
%the {\em average velocity}
%\begin{equation}
% V(r,t) = \int dots \int dv \, v \frac{\hat{\rho}(r,v,\dots,t)}{\rho(r,t)} \, ,
%\end{equation}
%and the velocity {\em variance}
%\begin{equation}
% \Theta(r,t) = \int \dots \int dv \, ( v - V(r,t) )^2
% \frac{\hat{\rho}(r,v,\dots,t)}{\rho(r,t)} \, .
%\end{equation}
To obtain an equation of this kind, %For this purpose, 
we will bring in the well-known fact that 
the temporal evolution of phase-space density $\hat{\rho}$ is given by the
{\em continuity equation} \cite{Keiz}
\begin{equation}
 \frac{\partial \hat{\rho}}{\partial t}
 + \nabla_{\!\vec{x}} \, \left( \hat{\rho} \frac{d\vec{x}}{dt} \right)
 = \left( \frac{\partial \hat{\rho}}{\partial t} \right)_{\rm tr}
\label{twenone}
\end{equation}
which again describes a conservation of the number of vehicles,
but this time in {\em phase-space} $\Omega = \{\mbox{all admissible states }
\vec{x} \}$. Whereas $\nabla_{\!\vec{x}} \, ( \hat{\rho} d\vec{x}/dt )$
reflects changes of phase-space density $\hat{\rho}$ 
due to a motion in phase space
$\Omega$ with velocity $d\vec{x}/dt$, the term $(\partial \hat{\rho} /\partial
t )_{\rm tr}$ delineates changes of $\hat{\rho}$ due to {\em discontinuous
transitions} between states.

\subsection{Prigogine's model}

In Prigogine's model the state $\vec{x}$ is given by the place $r$ and velocity
$v = dr/dt$ of a vehicle. The transition term 
$(\partial \hat{\rho}/\partial t)_{\rm tr}$
consists of a {\em relaxation term} $(\partial \hat{\rho}/\partial t)_{\rm rel}$
and an {\em interaction term} $(\partial \hat{\rho}/\partial t)_{\rm int}$
\cite{Andr,Pri,Prig}. 
Therefore, equation (\ref{twenone}) assumes the explicit form
\begin{equation}
 \frac{\partial \hat{\rho}}{\partial t} + \frac{\partial (\hat{\rho}v)}
 {\partial r} + \frac{\partial}{\partial v} \left( \hat{\rho} \frac{dv}{dt}
  \right) = \left( \frac{\partial \hat{\rho}}{\partial t} \right)_{\rm rel}
  + \left( \frac{\partial \hat{\rho}}{\partial t} \right)_{\rm int} \, .
\label{Prig1}
\end{equation} 
The interaction term $(\partial \rho/\partial t)_{\rm int}$ is intended
to describe the deceleration of vehicles to the velocity of the next car
ahead in situations when this moves slower and cannot be overtaken.
Prigogine \cite{Andr,Prig} suggests to describe processes of this kind by the
{\em Boltzmann equation}
\begin{mathletters}\label{Boltz}\begin{eqnarray}
 \left(\frac{\partial \hat{\rho}}{\partial t} \right)_{\rm int}
 &:=& \int\limits_v^\infty dw \, ( 1 - p ) 
 |v - w| \hat{\rho}(r,v,t) \hat{\rho}(r,w,t) 
\label{Boltza} \\
 &-&  \int\limits_0^v dw \, ( 1 - p ) |w - v| 
 \hat{\rho}(r,w,t) \hat{\rho}(r,v,t) 
\label{Boltzb} \\
&=& (1-p) \hat{\rho}(r,v,t) \int\limits_0^\infty dw \, (w-v) \hat{\rho}(r,w,t)
 \, . \nonumber
\end{eqnarray}\end{mathletters}
where $p$ denotes the 
%density- and maybe also velocity-dependent
probability that a slower car can be overtaken. Functional relations for
\begin{equation}
 p \equiv p(\rho,V,\Theta)
\end{equation}
are proposed in Refs. \cite{Prig,Phil,Hel3}.
The term (\ref{Boltza}) corresponds to situations where a vehicle with speed
$w > v$ must decelerate to speed $v$, causing an increase of phase-space
density $\hat{\rho}(r,v,t)$. The rate of these situations is proportional
\begin{itemize}
\item[1.] to the probability $(1-p)$ that passing is not possible
(which corresponds to the {\em `scattering cross section'} in kinetic
gas theory),
\item[2.] to the relative velocity $|v - w|$ of the interacting vehicles, 
\item[3.] to the phase-space density $\hat{\rho}(r,v,t)$ of vehicles which may
hinder a vehicle with velocity $w > v$, and 
\item[4.] to the phase-space
density $\hat{\rho}(r,w,t)$ of vehicles with velocity $w>v$ that may 
be affected by an interaction. 
\end{itemize}
Term (\ref{Boltzb}) describes a decrease
of phase-space density $\hat{\rho}(r,v,t)$ due to situations in which vehicles
with velocity $v$ must decelerate to a velocity $w < v$. 
A more detailed discussion of interaction term (\ref{Boltz}) can
be found in Refs. \cite{Prig,Pav}.
\par
Note that approach (\ref{Boltz}) assumes an instantaneous
adaptation of velocity which does not take any braking time. Moreover,
the deceleration process of the faster vehicle is assumed to happen
at the location $r$ of the slower vehicle, i.e. vehicles are
implicitly modelled as point-like objects without any space requirements.
The first
assumption is only justified for braking times that are short compared
to temporal changes of phase-space density $\hat{\rho}$, but modifications for
finite braking times are possible \cite{Hel3}.
The second assumption is only acceptable for very small densities 
at which the average headway distance is
much larger than average vehicle length plus safe distance. It will,
therefore, be corrected in Section VII. The corresponding modifications
also implicitly take into account the pair correlations of succeeding
vehicles \cite{Klim}. These are neglected by approach (\ref{Boltz}) 
due to its assumption of {\em `vehicular chaos'}, according to which the
velocities of vehicles are not correlated until they interact with each
other \cite{Pav,Nels}. 
\par
Now, we come to the description of acceleration processes by vehicles that do
not move with their desired speeds. In this connection, 
Prigogine suggests a {\em collective}
relaxation of the actual {\em velocity distribution}
\begin{equation}
 P(v;r,t) := \frac{\hat{\rho}(r,v,t)}{\rho(r,t)}
\end{equation}
towards an equilibrium velocity distribution $P_0(v)$ instead of an 
{\em individual} speed adjustment so that 
\begin{equation}
 \frac{dv}{dt} := 0 \, .
\end{equation}
In detail, Prigogine starts from the observation that free traffic is
characterized by a certain velocity distribution $P_0(v)$
which corresponds to the distribution 
$P_0(v_0)$ of {\em desired velocities} $v_0$. Moreover, he
assumes that the drivers' intention to get ahead with their desired speeds
causes the phase-space density $\hat{\rho}(r,v,t)$ to approach the
{\em equilibrium phase-space density}
\begin{equation}
 \hat{\rho}_0(r,v,t) := \rho(r,t) P_0(v)
\end{equation}
(exponentially) with a certain relaxation time $\tau$ which is given
by the average duration of acceleration processes. Therefore, 
Prigogine's relaxation term has the form \cite{Andr,Pri,Prig}
\begin{equation}
 \left(\frac{\partial \hat{\rho}}{\partial t} \right)_{\rm rel}
 := \frac{\rho(r,t)P_0(v) - \hat{\rho}(r,v,t)}{\tau} \, .
\label{relax}
\end{equation}
\par
Despite the merits of Prigogine's stimulating model,
this approach has been severely criticized \cite{Pav,Daganzo}. 
In a clear and detailed paper \cite{Pav}
Paveri-Fontana showed that Prigogine's model has a number of
peculiar properties which are not compatible with empirical findings.
For example, he demonstrates that the relaxation term (\ref{relax})
corresponds to {\em discontinuous} velocity changes which take place with a
certain, time-dependent rate. Furthermore, Daganzo criticized that,
according to (\ref{relax}), ``the desired speed distribution is a
property of the road and not the drivers'' \cite{Daganzo} which was already
noted by Paveri-Fontana \cite{Pav}. In reality, however, one can
distinguish different `personalities' of drivers: `aggressive' ones desire to 
drive faster, `timid' ones slower. Therefore, Paveri-Fontana \cite{Pav}
developed an improved gas-kinetic traffic model which corrects the
deficiencies of Prigogine's approach.

\subsection{Paveri-Fontana's model}

Paveri-Fontana assumes that each driver has an individual, characteristic
desired velocity $v_0$. Consequently, the associated states $\vec{x}$
are given by place $r$, velocity $v$, and desired velocity $v_0$ so
that Prigogine's phase-space density $\hat{\rho}(r,v,t)$ is replaced by
$\hat{\rho}(r,v,v_0,t)$. The corresponding gas-kinetic
equation (\ref{twenone}) explicitly reads\cite{Bemerk}
\begin{mathletters} \label{Paveri}
\begin{equation}
 \frac{\partial \hat{\rho}}{\partial t} + \frac{\partial (\hat{\rho} v)}
 {\partial r}  + \frac{\partial}{\partial v} \left( \hat{\rho}
 \frac{dv}{dt} \right) + \frac{\partial }{\partial v_0} \left(
 \hat{\rho} \frac{dv_0}{dt} \right) = \left( \frac{\partial \hat{\rho}}
 {\partial t} \right)_{\rm tr} \, .
\end{equation}
The term $\partial (\hat{\rho} dv_0/dt ) /\partial v_0$ can be neglected since
the desired velocity of each driver is normally time-independent during a trip
which implies
\begin{equation}
 \frac{dv_0}{dt} := 0 \, .
\end{equation}
In contrast to Prigogine, Paveri-Fontana describes the acceleration towards
the desired velocity $v_0$ by
\begin{equation}
 \frac{dv}{dt} := \frac{1}{\tau} ( v_0 - v )
\label{relaX}
\end{equation}
which means an {\em individual} instead of a {\em collective} relaxation. 
Relation (\ref{relaX}) can be easily replaced by other acceleration laws 
$dv/dt$
%like
%\begin{equation}
% \frac{dv}{dt} := \left\{
%\begin{array}{ll}
%a & \mbox{if } v < v_0 \\
%b & \mbox{if } v > v_0 \\
%0 & \mbox{otherwise}
%\end{array} \right.
%\end{equation}
or density-dependent driving programs as suggested by Alberti and Belli
\cite{Alb}. Alternatively, for acceleration processes an interaction approach 
can be formulated which was recently proposed by Nelson \cite{Nels}.
However, the assumption (\ref{relaX}) of exponential relaxation is a relatively
good approximation since drivers gradually reduce the acceleration
as they approach their desired velocity $v_0$.
\par
Paveri-Fontana needs the transition term $(\partial \hat{\rho}/\partial t
)_{\rm tr}$ only for the description of deceleration processes due to vehicular
interactions. For these he assumes the Boltzmann equation \cite{Pav}
\begin{eqnarray}
 \left(\frac{\partial \hat{\rho}}{\partial t} \right)_{\rm tr} 
 &:=& (1-p) \int\limits_v^\infty dw \int dw_0 \, |v - w| \hat{\rho}(r,v,w_0,t)
 \hat{\rho}(r,w,v_0,t) \nonumber \\
 &-& (1-p) \int\limits_0^v dw \int dw_0 \, |w - v| \hat{\rho}(r,w,w_0,t)
 \hat{\rho}(r,v,v_0,t) 
\label{intrel}
\end{eqnarray}\end{mathletters}
which has an analogous interpretation as (\ref{Boltz}). (For details cf.
Ref. \cite{Pav}.) Note that,
according to (\ref{intrel}), ``the velocity of the slow car
is unaffected by the interaction or by the fact of being passed'' \cite{Pav}
and that ``no driver changes his desired speed'' \cite{Pav} during
interactions.
Therefore, the interaction term (\ref{intrel}) fulfils the requirements called
for by Daganzo \cite{Daganzo}:
\begin{itemize}
\item[1.] that ``a car is an anisotropic particle that mostly responds to
frontal stimuli'' \cite{Daganzo} and that ``a slow car should be virtually
unaffected by its interaction with faster cars passing it (or queueing behind
it)'' \cite{Daganzo}. 
\item[2.] that ``interactions do not change the `personality'
(aggressive/timid) of any car'' \cite{Daganzo}.
\end{itemize}
Finally, note that the proportion of vehicles jamming behind slower
cars cannot accelerate. This circumstance can be taken into account by
a density- and maybe velocity- or variance-dependence of the relaxation time
\cite{Prig,Phil,Hel3}:
\begin{equation}
 \tau \equiv \tau(\rho,V,\Theta) \, .
\end{equation}
In order to compare  Paveri-Fontana's traffic equation with Prigogine's one we
integrate equation (\ref{Paveri}) with respect to $v_0$ and obtain
the {\em reduced Paveri-Fontana equation}
\begin{eqnarray}
& & \frac{\partial \tilde{\rho}}{\partial t} + \frac{\partial
 (v \tilde{\rho})}{\partial r} + \frac{\partial}{\partial v}
 \left[ \tilde{\rho}(r,v,t) \frac{\tilde{V}_0(v;r,t) - v}{\tau} \right] 
\nonumber \\
&=& (1-p) \tilde{\rho}(r,v,t) \int\limits_0^\infty 
 dw \, (w - v) \tilde{\rho}(r,w,t)  \, .
\label{Fontana}
\end{eqnarray}                                           
Here, we have introduced the {\em reduced phase-space density}
\begin{equation}
 \tilde{\rho}(r,v,t) := \int dv_0 \,  \hat{\rho}(r,v,v_0,t)
\end{equation}
and the quantity
\begin{equation}
 \tilde{V}_0(v;r,t) := \int dv_0 \, v_0 \frac{\hat{\rho}(r,v,v_0,t)}
 {\tilde{\rho}(r,v,t)} \, . 
\end{equation}
The only difference with respect to Prigogine's formulation (\ref{Prig1})
to (\ref{relax}) is obviously the other relaxation term.

\section{Derivation of macroscopic traffic equations}

Since we are mainly interested in the temporal evolution of collective
(`macroscopic') quantities like the spatial {\em density}
\begin{equation}
 \rho(r,t) := \int dv \, \tilde{\rho}(r,v,t) 
\end{equation}
per lane, the {\em average velocity}
\begin{equation}
 V(r,t) \equiv \langle v \rangle := \int dv \, v
 \frac{\tilde{\rho}(r,v,t)}{\rho(r,t)} \, ,
\end{equation}
and the velocity {\em variance}
\begin{eqnarray}
 \Theta(r,t) \equiv \langle [v -V(r,t)]^2 \rangle 
 &:=& \int dv \, [v - V(r,t)]^2
 \frac{\tilde{\rho}(r,v,t)}{\rho(r,t)}  \nonumber \\
% &=& \int dv \, v^2
% \frac{\tilde{\rho}(r,v,t)}{\rho(r,t)} - V^2 
 &=& \langle v^2 \rangle - [V(r,t)]^2 
\end{eqnarray}
we will now derive equations for the moments $m_{k,0}$ with
\begin{equation}                               
  m_{k,l}(r,t) \equiv \rho(r,t) \langle v^k (v_0)^l \rangle 
 := \int dv \int dv_0 \, v^k (v_0)^l \hat{\rho}(r,v,v_0,t) \, .
\end{equation}
By multiplying Paveri-Fontana's equation (\ref{Fontana}) 
with $v^k$ and integrating
with respect to $v$ we obtain \cite{Pav}, via partial integration,
\begin{mathletters}\label{moment1}
\begin{eqnarray}
& & \frac{\partial}{\partial t} m_{k,0} + \frac{\partial}{\partial r}
 m_{k+1,0} + \int dv \, v^k \frac{\partial}{\partial v}
 \left( \tilde{\rho} \frac{\tilde{V}_0(v) - v}{\tau} \right) 
\nonumber \\
&=& \frac{\partial}{\partial t} m_{k,0} + \frac{\partial}{\partial r}
 m_{k+1,0} - \int dv \, k v^{k-1} 
 \left( \tilde{\rho} \frac{\tilde{V}_0(v) - v}{\tau} \right) 
\nonumber \\
&=& \frac{\partial}{\partial t} m_{k,0} + \frac{\partial}{\partial r}
 m_{k+1,0} - \frac{k}{\tau} (m_{k-1,1} - m_{k,0}) \\
&=& (1-p) \int dv \, \tilde{\rho}(r,v,t) \int dw \,
 (w v^k - v^{k+1}) \tilde{\rho}(r,w,t) 
\nonumber \\
&=& (1-p) (m_{1,0}m_{k,0} - m_{k+1,0}m_{0,0}) \, .
\end{eqnarray}\end{mathletters}
Applying the analogous procedure to Prigogine's model (\ref{Prig1}) 
to (\ref{relax}), for the moments
\begin{equation}
  m_{k,0}(r,t) \equiv \rho(r,t) \langle v^k \rangle 
 := \int dv \, v^k \hat{\rho}(r,v,t)
\end{equation}
one can derive the equations 
\begin{eqnarray}
 \frac{\partial}{\partial t} m_{k,0} &+& \frac{\partial}{\partial r}
 m_{k+1,0} = \frac{1}{\tau} (m_{0,k} - m_{k,0} ) 
\nonumber \\
&+& (1-p) (m_{1,0}m_{k,0} - m_{k+1,0} m_{0,0} ) \label{moment2}
\end{eqnarray}
(cf. \cite{Pav}) where 
\begin{eqnarray}
  m_{0,k}(r,t) &:=& \int dv_0 \, (v_0)^k \hat{\rho}_0(r,v_0,t) \nonumber \\
 &=& \rho(r,t) \int dv_0 \, (v_0)^k P_0(v_0) \, .
\end{eqnarray}
A comparison of moment equations (\ref{moment1}) with (\ref{moment2}) shows 
that Prigogine's and Paveri-Fontana's model lead to identical equations
for spatial density $\rho(r,t) = m_{0,0}(r,t)$ and average
velocity $V(r,t) = m_{1,0}(r,t)/\rho(r,t)$, despite the different approaches
for the relaxation term. However, the equations for higher order moments
$m_{k,0}(r,t)$ with $k \ge 2$ differ.
\par
Obviously, equations (\ref{moment1}) as well as 
(\ref{moment2}) represent a hierarchy
of non-closed equations since the equation for the $k$th moment $m_{k,0}$
depends on the $(k+1)$st moment $m_{k+1,0}$. As a consequence, the
{\em density equation}
\begin{equation}
 \frac{\partial \rho}{\partial t} + \frac{\partial (\rho V)}{\partial r}
 = 0
\label{dens}
\end{equation}
depends on average velocity $V$, the {\em velocity equation}
\begin{eqnarray}
 \frac{\partial V}{\partial t} + V \frac{\partial V}{\partial r}
 &=& - \frac{1}{\rho} \frac{\partial (\rho\Theta)}{\partial r}
 + \frac{1}{\tau} ( V_0 - V ) - (1-p) \rho \Theta \nonumber \\
 &=& - \frac{1}{\rho} \frac{\partial {\cal P}}{\partial r}
 + \frac{1}{\tau} [ V_e(\rho,V,\Theta) - V ] 
\label{veloc}
\end{eqnarray}
on variance $\Theta$, etc. Here, we have introduced the
{\em average desired velocity}
\begin{equation}
 V_0(r,t) := \int dv \int dv_0 \, v_0 \frac{\hat{\rho}(r,v,v_0,t)}{\rho(r,t)}
\, ,
\end{equation}
the so-called {\em `traffic pressure'} \cite{Pri,Phil,Phil2}
\begin{eqnarray}
 {\cal P}(r,t) &:=& \frac{1}{\rho(r,t)} \int dv \, (v - V) \tilde{\rho}(r,v,t)
 \int dw \, (v-w) \tilde{\rho}(r,w,t) \nonumber \\
 &=& \int dv \, (v-V)^2 \tilde{\rho}(r,v,t) = \rho(r,t) \Theta(r,t) \, ,
\label{pr}
\end{eqnarray}
and the {\em equilibrium velocity}
\begin{equation}
 V_e(\rho,V,\Theta) := V_0 - 
 \tau(\rho,V,\Theta) [1 - p(\rho,V,\Theta)] {\cal P}
\label{vauE}
\end{equation}
which is related with stationary and spatially homogeneous traffic flow. 
\par
Equations (\ref{dens}) and (\ref{veloc}) are
easily derivable from the moment equations (\ref{moment1}) and (\ref{moment2})
respectively by use of $m_{0,0} = \rho$ and
\begin{equation}                                                    
 \frac{\partial m_{1,0}}{\partial t} = \frac{\partial (\rho V)}{\partial t}
 = \rho \frac{\partial V}{\partial t} + V \frac{\partial \rho}{\partial t}
 \, .
\end{equation}
The {\em variance equation} is obtained analogously. For the traffic
equation of Paveri-Fontana 
%which is superior to that of Prigogine (cf. the discussion in Section ...) 
it reads
\begin{eqnarray}
 \frac{\partial \Theta}{\partial t} + V\frac{\partial \Theta}{\partial r}
&=& - 2\Theta \frac{\partial V}{\partial r} - \frac{1}{\rho}
 \frac{\partial (\rho \Gamma)}{\partial r} \nonumber \\
&+& \frac{2}{\tau} ( {\cal C} - \Theta ) - (1-p) \rho \Gamma \nonumber \\
&=& - \frac{2{\cal P}}{\rho} \frac{\partial V}{\partial r} - \frac{1}{\rho}
 \frac{\partial {\cal J}}{\partial r} \nonumber \\
&+& \frac{2}{\tau} [\Theta_e(\rho,V,\Theta,{\cal C},{\cal J}) - \Theta ] 
\label{var}
\end{eqnarray}
and depends on the {\em covariance}
\begin{eqnarray}
 {\cal C}(r,t) &\equiv & \langle (v - V)(v_0 - V_0) \rangle \nonumber \\
 &:=& \int dv_0 \int dv \, (v - V)(v_0 - V_0) \frac{\hat{\rho}(r,v,v_0,t)}
 {\rho(r,t)} \nonumber \\
 &=& \int dv \, (v - V) [\tilde{V}_0(v) - V_0]
 \frac{\tilde{\rho}(r,v,t)}{\rho(r,t)}
\label{co}
\end{eqnarray}
as well as the {\em third central moment}
\begin{equation}
 \Gamma(r,t) \equiv \langle (v - V)^3 \rangle := \int dv \, (v - V)^3 
 \frac{\tilde{\rho}(r,v,t)}{\rho(r,t)} \, .
\end{equation}
In addition, we have introduced the {\em flux density of velocity variance} 
\begin{eqnarray}
 {\cal J}(r,t) &:=& 
 \frac{1}{\rho(r,t)} \int dv \, (v - V)^2 \tilde{\rho}(r,v,t)
 \int dw \, (v-w) \tilde{\rho}(r,w,t) \nonumber \\
 &=& \int dv \, (v-V)^3 \tilde{\rho}(r,v,t) = \rho(r,t) \Gamma(r,t)
\end{eqnarray}
(which corresponds to the `heat flow' in conventional fluid-dynamics)
and the {\em equilbrium variance}
\begin{equation}
 \Theta_e(\rho,V,\Theta,{\cal C},{\cal J}) := {\cal C} -
 \frac{\tau(\rho,V,\Theta)}{2} [1 - p(\rho,V,\Theta)] 
 {\cal J} \, .
\label{thetaE}
\end{equation}

\subsection{Approximate closed macroscopic traffic equations}

We will now face the problem of closing the hierarchy of moment equations by a
suitable approximation. The simplest approximations replace a macroscopic
quantity $Q(r,t)$ (which would be determined by a dynamic equation) by its
{\em equilibrium value} $Q_{e}$ which belongs to the stationary and
spatially homogeneous solution. Approximations of this kind are
zeroth-order approximations. 
The simplest one is obtained by a substitution
of $V(r,t)$ (which actually obeys Eq. (\ref{veloc}))
by the {\em equilibrium velocity}
\begin{equation}
 V_{e}(\rho) := V_0 - \tau(\rho) [1-p(\rho)] \rho \Theta_{e}(\rho) 
\label{vrela}
\end{equation}
(cf. (\ref{vauE})). 
Equations (\ref{dens}), (\ref{vrela}) obviously correspond
to the model (\ref{cont}), (\ref{LW}) of Lighthill and Whitham. 
Relation (\ref{vrela}) specifies the equilibrium velocity-density relation
(\ref{LW}) in accordance with Paveri-Fontana's traffic equation. It could
be interpreted as a theoretical result concerning the dependence of
$V_{e}(\rho)$ on the microscopic processes of traffic flow: According to
(\ref{vrela}), the equilibrium velocity $V_{e}(\rho)$ is given by the
average desired velocity $V_0$
diminished by a term arising from necessary deceleration maneuvers due to
interactions of vehicles.
\par
However, according to equation (\ref{veloc}), the approximation 
$V(r,t) \approx V_{e}[\rho(r,t)]$
is only justified for $\tau \rightarrow 0$ which
is not compatible with empirical data. Consequently, the latter
does not adequately describe non-equilibrium situations like on-ramp traffic
or stop-and-go traffic where the velocity is not uniquely given by the spatial
density $\rho(r,t)$.
\par
Another zeroth-order approximation is found by leaving Eq. (\ref{veloc})
unchanged but replacing the dynamic variance
$\Theta(r,t)$ by the equilibrium variance
\begin{equation}
 \Theta_{e}(\rho,V) := {\cal C}_{e}(\rho,V) - \frac{\tau(\rho,V)}{2} 
 [1 - p(\rho,V)] \rho \Gamma_{e}(\rho,V) 
\label{thet}
\end{equation}
(cf. (\ref{thetaE})). (Here, the subscript $e$ shall again indicate the 
equilibrium-value or -relation of a function.) 
The resulting model (\ref{dens}), (\ref{veloc}), (\ref{thet}) obviously 
corresponds to the model (\ref{cont}), (\ref{Phillips}) of 
Phillips, this time specifying the equilibrium variance-density relation in 
accordance with Paveri-Fontana's traffic model. A complete agreement
between (\ref{thet}) and (\ref{Philb}) results for 
${\cal C}_e(\rho,V) \equiv {\cal C}_e(\rho)$, $\Gamma_e(\rho,V) \equiv
\Gamma_e(\rho)$, and a special choice of the
functional relation 
$\tau(\rho,V) [1 - p(\rho,V)] \equiv \tau(\rho) [1 - p(\rho)]$.
\par
However, it is not fully justified to assume that the variance
$\Theta(r,t)$ is always in equilibrium $\Theta_{e}(\rho,V)$, since the
corresponding relaxation time $2/\tau$ is of the order of the relaxation time
$1/\tau$ for the velocity $V(r,t)$. Moreover, the approximation 
$\Theta(r,t) \approx \Theta_{e}[\rho(r,t),V(r,t)]$ does not describe the
empirically observed increase of variance $\Theta$ directly before 
a traffic jam develops \cite{Kuehne,Hel1}. Therefore, we also need the
dynamic variance equation (\ref{var}). The remaining problem 
is how to obtain suitable relations for $\Gamma(r,t)$ and ${\cal C}(r,t)$.

\subsection{Euler-like traffic equations}

Before looking for dynamic relations for $\Gamma(r,t)$ and ${\cal C}(r,t)$,
it is plausible first to look for equilibrium relations which apply to
stationary and spatially homogeneous traffic. For this purpose
we require the equilibrium
solution $\hat{\rho}_e(v,v_0)$ of Paveri-Fontana's traffic equation
(\ref{Paveri}). 
\par
Unfortunately, it seems impossible to find an analytical expression for
$\hat{\rho}_e(v,v_0)$, but in order to derive
equations for the velocity moments
$\langle v^k \rangle$ we are mainly interested in, it is sufficient to
find the stationary and spatially homogeneous solution
$\tilde{\rho}_e(v)$ of the {\em reduced} Paveri-Fontana equation (\ref{Fontana}).
For this we need to know the relation
\begin{equation}
 \tilde{V}_0(v) = a_0 + a_1 \, \delta v + a_2 \, (\delta v)^2 + \dots +
 a_n \, (\delta v)^n 
\label{expansion}
\end{equation}
with 
\begin{equation}
 \delta v := v - V
\end{equation}
and arbitrary $n$. However, the equation that determines $\tilde{V}_0(v)$
depends on the unknown quantity
\begin{equation}
 \tilde{\Theta}_0(v) := \int dv_0 \, (v_0 - V_0)^2
 \frac{\hat{\rho}_e(v,v_0)}{\tilde{\rho}_e(v)} 
\end{equation}
etc. so that we are again confronted with a non-closed hierarchy of equations.
\par
Luckily, from empirical data and microsimulations 
we know that the equilibrium velocity-distribution 
\begin{equation}
 P_e(v) := \frac{\tilde{\rho}_e(v)}{\rho_e}
\end{equation}
(at least in the range 
of {\em stable} traffic without {\em stop-and-go waves}) 
is approximately a Gaussian distribution \cite{Pampel,MuPi,Phil,Notiz,MIC1}:
\begin{equation}
 P_e(v) = {\textstyle\frac{1}{\sqrt{2 \pi \Theta_e}}} \,
 \mbox{e}^{-(v - V_e)^2/(2\Theta_e)} \, . \label{Gaussian}
\end{equation}
Inserting (\ref{expansion}) and (\ref{Gaussian}) into the equation
\begin{equation}
 \frac{\partial}{\partial v} \left( \tilde{\rho}_e \frac{\tilde{V}_0(v) -
   v}{\tau} \right) = - (1-p) \tilde{\rho}_e \rho_e \, \delta v
\end{equation}
which corresponds to equation (\ref{Fontana}) in the stationary and spatially
homogeneous case, we find the condition
\begin{mathletters}\begin{eqnarray}
& &  \frac{\partial}{\partial v} \left( \tilde{\rho}_e \frac{\tilde{V}_0(v) -
   v}{\tau} \right)
= \frac{\tilde{V}_0(v) - v}{\tau} \frac{\partial \tilde{\rho}_e }{\partial v}
+ \frac{\tilde{\rho}_e}{\tau} \left( \frac{\partial \tilde{V}_0(v)}{\partial
  v} - 1 \right) \nonumber \\
&=& \frac{\tilde{\rho}_e}{\tau} \left[ (a_1 - 1) +
 \left( 2 a_2 - \frac{a_0 - V_e}{\Theta_e} \right) \, \delta v
 + \left( 3 a_3 - \frac{a_1 - 1}{\Theta_e} \right) \, (\delta v)^2 \right.
 \nonumber \\
 & & \dots + \left. 
 \left( k a_k - \frac{a_{k-2}}{\Theta_e} \right) \, (\delta v)^{k-1}
 \dots - \frac{a_{n-1}}{\Theta_e} \, (\delta v)^n - \frac{a_n}{\Theta_e} \, 
 (\delta v)^{n+1} \right] \label{Konda} \\
&\stackrel{!}{=}& - (1-p) \tilde{\rho}_e \rho_e \, \delta v \, .
\label{Kondb}
\end{eqnarray}\end{mathletters}
A comparison of the coefficients of $(\delta v)^k$ in (\ref{Konda}) and
(\ref{Kondb}) leads to 
\begin{equation}
 a_n = 0 \, , \qquad a_{n-1} = 0 \, , \quad \dots \quad a_2 = 0 \, ,
 \qquad a_1 = 1 \, ,
\end{equation}
and
\begin{equation}
 a_0 = V_e + \tau (1-p) \rho_e \Theta_e = V_0 \, ,
\end{equation}
where we have utilized relation (\ref{vauE}) with (\ref{pr}). Consequently,
%the empirically justified 
for equilibrium situations velocity distribution (\ref{Gaussian}) implies
\begin{equation}
 \tilde{V}_0(v) = V_0 + \delta v \, .
\label{vau0vau}
\end{equation}
\par
With (\ref{Gaussian}) and (\ref{vau0vau}) we can now derive equilibrium
relations for ${\cal C}$ and $\Gamma$. One obtains
\begin{equation}
 \Gamma_e = 0
\end{equation}
and
\begin{equation}
 {\cal C}_e = \Theta_e \, .
\label{one}
\end{equation}
\par
Next, we are looking for relations for non-equilibrium cases.
Assuming that the {\em velocity distribution}
\begin{equation}
 P(v;r,t) := \frac{\tilde{\rho}(r,v,t)}{\rho(r,t)}
\end{equation}
locally approaches the equilibrium distribution 
$P_e[V(r,t),\Theta(r,t)]$ very 
rapidly, we can apply the zeroth-order
{\em approximation of local equilibrium:}
\begin{eqnarray}
 P(v;r,t) &\approx& P_{(0)}(v;r,t) := P_e[V(r,t),\Theta(r,t)] \nonumber \\
 &=& {\textstyle\frac{1}{\sqrt{2\pi \Theta(r,t)}}} \,
 \mbox{e}^{-[v - V(r,t)]^2/[2\Theta(r,t)]} \, .
\label{local}
\end{eqnarray}
Furthermore, in order to fulfil the compatibility condition
\begin{equation}
 {\cal C}(r,t) = \int dv [v - V(r,t)][v_0 - \tilde{V}_0(v;r,t)]
 P(v;r,t)
\end{equation}
(cf. (\ref{co})), we must generalize relation (\ref{vau0vau}) to
\begin{equation}
 \tilde{V}_0(v;r,t) = V_0 + \frac{{\cal C}(r,t)}{\Theta(r,t)}\, \delta v
\label{missing}
\end{equation}
which is fully consistent with (\ref{one}).
Relations (\ref{local}) and (\ref{missing}) yield zeroth-order relations for
the spatio-temporal variation of ${\cal C}(r,t)$ and ${\cal J}(r,t)$:
For the flux density of velocity variance we find
\begin{equation}
 {\cal J}(r,t) \approx {\cal J}_{(0)}(\rho,V,\Theta) 
 = \rho \Gamma_{(0)}(\rho,V,\Theta) = 0 \, ,
\end{equation}
whereas for the covariance the dynamic equation 
\begin{equation}
 \frac{\partial {\cal C}}{\partial t} + V \frac{\partial {\cal C}}{\partial r}
 = - {\cal C} \frac{\partial V}{\partial r} - \frac{{\cal P}}{\rho} 
 \frac{\partial V_0}{\partial r} + \frac{1}{\tau}(\Theta_0
  - {\cal C}) - 2 (1-p) \rho {\cal C} \sqrt{\frac{\Theta}{\pi}} 
\label{Covar}
\end{equation}
can be derived from the reduced Paveri-Fontana equation (\ref{Fontana}) due to
\begin{eqnarray}
 & & \int dv \int dv_0 \, (\delta v)^2 \delta v_0 \hat{\rho}(r,v,v_0,t)
\nonumber \\
 &=& \int dv \, (\delta v)^2 [ \tilde{V}_0(v) - V_0 ] \tilde{\rho}(r,v,t)
 \nonumber \\
 &=&  \int dv \, (\delta v)^3 \frac{{\cal C}}{\Theta} \tilde{\rho}(r,v,t)
 = {\cal J} \frac{{\cal C}}{\Theta} 
\label{diesda}
\end{eqnarray}
($\delta v_0 := v_0 - V_0$). (The somewhat lengthy but straightforward
calculation is presented in Ref. \cite{hifi}.) 
\par
In the zeroth-order {\em covariance equation} (\ref{Covar}) the quantity
\begin{equation}
 \Theta_0(r,t) := \int dv \int dv_0 \, [v_0 - V_0(r,t)]^2
 \frac{\hat{\rho}(r,v,v_0,t)}{\rho(r,t)}
\end{equation}
denotes the {\em variance of desired velocities}. The term $-\Theta \partial
V_0/\partial r$ normally vanishes since the average desired velocity $V_0$
is approximately constant almost everywhere (cf. \cite{Bemerk}).
Due to (\ref{one}),
the {\em equilibrium variance} related to stationary and homogeneous traffic
is obviously determined by the implicit relation
\begin{equation}
 \Theta_{e}(\rho_e,V_e,\Theta_e) = {\cal C}_{e}(\rho_e,V_e,\Theta_e) 
 = \Theta_0 - 2 \tau (1-p) \rho_e \Theta_e \sqrt{\frac{\Theta_e}{\pi}} \, .
\label{thet1}
\end{equation}
\par
Inserting the above results into equations 
(\ref{dens}), (\ref{veloc}), and (\ref{var}),
we obtain the following zeroth-order approximations of the density-,
velocity-, and variance-equation respectively:
\begin{equation}
 \frac{\partial \rho}{\partial t} + V\frac{\partial \rho}{\partial r}
 = - \rho \frac{\partial V}{\partial r} \, , 
\label{rho}
\end{equation}
\begin{eqnarray}
 \frac{\partial V}{\partial t} + V\frac{\partial V}{\partial r}
 &=& - \frac{1}{\rho} \frac{\partial (\rho \Theta)}{\partial r}
 + \frac{1}{\tau} (V_0 - V) - (1-p) \rho \Theta \nonumber \\
 &=& - \frac{1}{\rho} \frac{\partial {\cal P}}{\partial r}
 + \frac{1}{\tau} [V_e(\rho,V,\Theta) - V] \, ,
\label{vau}
\end{eqnarray}
\begin{eqnarray}
 \frac{\partial \Theta}{\partial t} + V\frac{\partial \Theta}{\partial r}
 &=& - 2 \Theta \frac{\partial V}{\partial r}  %- \frac{1}{\rho}
% \frac{\partial (\rho \Gamma_{(0)})}{\partial r} 
+ \frac{2}{\tau}
 ( {\cal C} - \Theta )  \nonumber \\
% & & - (1-p)\rho \Gamma_{(0)}(\rho,V,\Theta)
% \nonumber \\
 &=& - \frac{2 {\cal P}}{\rho} \frac{\partial V}{\partial r} % - \frac{1}{\rho}
 % \frac{\partial {\cal J}_{(0)}}{\partial r} 
 + \frac{2}{\tau} ( {\cal C} - \Theta ) \, .
\label{theta}
\end{eqnarray}
\par
Equations (\ref{rho}), (\ref{vau}), and (\ref{theta}) are the 
{\em `Euler-like equations'} of vehicular 
traffic \cite{Rem2}. In comparison with the
Euler equations for ordinary fluids \cite{Jaeckle,Rieck,Huang,Lib} they 
contain additional terms:
\begin{itemize}
\item[1.] The terms $(V_0 - V)/\tau$ and $2({\cal C} - \Theta)/\tau$ arise
from the acceleration of vehicles towards the drivers' desired velocities
$v_0$, i.e. they are a consequence of the fact that driver-vehicles units are 
{\em active} systems.
\item[2.] The term $-(1-p)\rho\Theta$ 
results from the vehicles' interactions. It would vanish if momentum 
would be a {\em collisional invariant} during vehicular interactions like
this is the case for atomic collisions \cite{Keiz}. However,
without this term the `vehicular fluid' would speed up at bottlenecks which
is, of course, unrealistic.
\end{itemize}
Moreover, the covariance equation (\ref{Covar}) is a complementary equation
which arises from the drivers' tendency to move with their desired velocities
$v_0$. %It is an effect of the circumstance that driver-vehicle units
%are {\em active} systems.

\subsection{Equilibrium relations and fundamental diagram}

For vehicular traffic, the only dynamic quantity 
that remains unchanged in a closed
system (i.e. a circular road) is the average spatial density $\bar{\rho}$
(due to the conservation of the number of vehicles). As a consequence, the
equilibrium traffic situation is uniquely determined by $\bar{\rho}$ which
obviously agrees with the equilibrium density $\rho_e$. Equilibrium relations
for the average velocity $V_e(\rho_e)$ and the velocity variance
$\Theta_e(\rho_e)$ in dependence of $\rho_e = \bar{\rho}$ can be obtained from
equations (\ref{vauE}) and (\ref{thet1}) if the relations 
$p(\rho,V,\Theta)$ and $\tau(\rho,V,\Theta)$ are given 
(cf. \cite{Prig,Phil}). A simple procedure for finding a solution of these
implicit equations is to numerically integrate the equations
\begin{eqnarray}
 \frac{dV}{dy} &=& V_e[\rho_e,V(y),\Theta(y)] - V(y) \nonumber \\
&=& V_0 - \tau(\rho_e,V,\Theta)[1-p(\rho_e,V,\Theta)] \varrho_e \Theta - V 
 \, , \label{VR} \\
\frac{d\Theta}{dy} &=& \Theta_e[\rho_e,V(y),\Theta(y)] - \Theta(y) \nonumber \\
&=& \Theta_0 - 2 \tau(\rho_e,V,\Theta) [1-p(\rho_e,V,\Theta)] \varrho_e 
 \Theta \sqrt{\frac{\Theta}{\pi}} - \Theta \label{TR}
\end{eqnarray}
until $dV/dy = 0$ and $d\Theta/dy = 0$. 
Here, we have replaced $\rho_e$ by $\varrho_e = \varrho_e(\rho_e,V)$
in accordance with section VII B in order to take into account
the space requirements of vehicles. The theoretical results for the
{\em equilibrium velocity-density relation} $V_e(\rho_e) =
\displaystyle \lim_{y \rightarrow \infty} V(y)$, the
{\em equilibrium variance-density relation} $\Theta_e(\rho_e) =
\displaystyle \lim_{y \rightarrow \infty} \Theta(y)$, and the
{\em fundamental diagram}
\begin{equation}
 q_e(\rho_e) := \rho_e V_e(\rho_e)
\label{Fund}
\end{equation}
can be directly compared with empirical data.
\par
If, however, $p(\rho,V,\Theta)$ or $\tau(\rho,V,\Theta)$ are unknown relations,
it is still possible to derive from the fundamental diagram $q_e(\rho_e)$
the equilibrium variance-density relation $\Theta_e(\rho_e)$
for which an empirical relation seems
to be missing: From (\ref{VR}) and (\ref{Fund}) we get
\begin{equation}
 \tau (1-p) \varrho_e \Theta_e(\rho_e) = V_0 - V_e(\rho_e) = V_0
 - \frac{q_e(\rho_e)}{\rho_e} \, .
\label{Eins}
\end{equation}
Inserting this into (\ref{thet1}) we find
\begin{eqnarray}
 \Theta_e(\rho_e) &=& \Theta_0 - 2 \tau (1-p) \varrho_e \Theta_e(\rho_e)
 \sqrt{\frac{\Theta_e(\rho_e)}{\pi}} \nonumber \\
 &=& \Theta_0 - 2 [ V_0 - V_e(\rho_e) ] \sqrt{\frac{\Theta_e(\rho_e)}{\pi}} \, .
\end{eqnarray} 
This results in a quadratic equation for the {\em standard deviation 
$\sqrt{\Theta_e(\rho_e)}$ of vehicle velocities} which is solved by
\begin{equation}
 \sqrt{\Theta_e(\rho_e)} = - \frac{V_0 - V_e(\rho_e)}{\sqrt{\pi}}
 + \sqrt{\frac{[V_0 - V_e(\rho_e)]^2}{\pi} + \Theta_0 } \, .
\label{Zwei}
\end{equation}

\section{Approximate Solution of Paveri-Fontana's traffic equation}

The traffic equation of Paveri-Fontana was mathematically investigated in
several papers dealing with the existence, uniqueness, and numerical 
determination of a solution which satisfies 
the non-linear initial-value boundary problem \cite{PAV1,PAV2,PAV3}.
However, %as far as the author of this paper knows, 
the approximate dynamic solution of the reduced Paveri-Fontana equation
(\ref{Fontana}) which will be presented in this section
has not been proposed before.
\par
As one would expect, in non-equilibrium situations
the zeroth-order approximation (\ref{local}) 
does not solve the reduced Paveri-Fontana equation
(\ref{Fontana}) exactly. Therefore, we write
\begin{equation}
 \tilde{\rho}(r,v,t) =: \tilde{\rho}_{(0)}(r,v,t)
 + \tilde{\rho}_{(1)}(r,v,t)
\end{equation}
with
\begin{equation}
 \tilde{\rho}_{(0)}(r,v,t) := \rho(r,t) P_{(0)}(v;r,t)
 = {\textstyle\frac{\rho(r,t)}{\sqrt{2\pi \Theta(r,t)}}} \,
 \mbox{e}^{-[v - V(r,t)]^2/[2\Theta(r,t)]}
\end{equation}
and try to derive a relation for the deviation
$\tilde{\rho}_{(1)}(r,v,t)$. Utilizing that the correction term 
$\tilde{\rho}_{(1)}(r,v,t)$ will usually be small compared to 
$\tilde{\rho}_{(0)}(r,v,t)$ we have
\begin{equation}
 \tilde{\rho}_{(1)}(r,v,t) \ll \tilde{\rho}_{(0)}(r,v,t)
\label{ll}
\end{equation}
and get 
\begin{eqnarray}
& &  \frac{\partial \tilde{\rho}}{\partial t}
 + v \frac{\partial \tilde{\rho}}{\partial r} 
 + \frac{\partial}{\partial v} \left(
 \tilde{\rho} \frac{\tilde{V}_0(v) - v}{\tau}  \right)
\nonumber \\
&\approx& \frac{\partial \tilde{\rho}_{(0)}}{\partial t}
 + v \frac{\partial \tilde{\rho}_{(0)}}{\partial r}
 + \frac{\partial}{\partial v} \left(
 \tilde{\rho}_{(0)} \frac{\tilde{V}_0(v) - v}{\tau} \right) \nonumber \\
 &=& \frac{\partial \tilde{\rho}_{(0)}}{\partial t}
 + v \frac{\partial \tilde{\rho}_{(0)}}{\partial r}
 + \frac{\tilde{V}_0(v) - v}{\tau} \frac{\partial \tilde{\rho}_{(0)}}
 {\partial v} + \frac{\tilde{\rho}_{(0)}}{\tau}
 \left( \frac{\partial \tilde{V}_0(v)}{\partial v} - 1 \right) \, .
\label{num1}
\end{eqnarray}
(For a detailled discussion of this approximation 
cf. \cite{Jaeckle,Rieck,Lib}.)
Now, introducing the abbreviation
\begin{equation}
 \frac{d}{dt} := \frac{\partial}{\partial t} + v \frac{\partial}{\partial r}
\end{equation}
we can write
\begin{eqnarray}
 \frac{\partial \tilde{\rho}_{(0)}}{\partial t} + v \frac{\partial \tilde{\rho}_{(0)}}
 {\partial r} &=& \frac{d\tilde{\rho}_{(0)}}{dt} 
 = \frac{\partial \tilde{\rho}_{(0)}}{\partial \rho}\frac{d\rho}{dt}
 + \frac{\partial \tilde{\rho}_{(0)}}{\partial V}\frac{dV}{dt}
 + \frac{\partial \tilde{\rho}_{(0)}}{\partial \Theta}\frac{d\Theta}{dt}\nonumber \\
&=& \frac{\tilde{\rho}_{(0)}}{\rho}\frac{d\rho}{dt} 
 + \frac{\tilde{\rho}_{(0)}}{\Theta} \delta v \frac{dV}{dt}
 + \frac{\tilde{\rho}_{(0)}}{2\Theta} \left( \frac{(\delta v)^2}{\Theta} - 1
   \right) \frac{d\Theta}{dt} \, .
\label{num2}
\end{eqnarray}
%with
%\begin{equation}
% \delta v := v - V \, .
%\end{equation}
Relations for $d\rho/dt$, $dV/dt$, and $d\Theta/dt$ can be obtained
from the Euler-like equations (\ref{rho}), (\ref{vau}), and (\ref{theta})
via
\begin{equation}
 \frac{d}{dt} = \frac{\partial}{\partial t} + V \frac{\partial}{\partial r}
 + \delta v \frac{\partial}{\partial r} \, .
\end{equation}
We find
\begin{mathletters}\label{num3}
\begin{equation}
 \frac{d\rho}{dt} = \delta v \frac{\partial \rho}{\partial r} - \rho
 \frac{\partial V}{\partial r} \, ,
\end{equation}
\begin{equation}
 \frac{dV}{dt} = \delta v \frac{\partial V}{\partial r} - \frac{1}{\rho}
 \frac{\partial (\rho\Theta)}{\partial r} + \frac{1}{\tau} [ V_e(\rho,V,\Theta)
 - V ] \, ,
\end{equation}
and
\begin{equation}
 \frac{d\Theta}{dt} = \delta v \frac{\partial \Theta}{\partial r}
 - 2\Theta \frac{\partial V}{\partial r}
 + \frac{2}{\tau} ( {\cal C} - \Theta ) \, .
\end{equation}\end{mathletters}
For the interaction term we apply a linear approximation in 
$\tilde{\rho}_{(1)}(r,v,t)$ which is justified by relation (\ref{ll}). The result is
\begin{mathletters}\label{num4}
\begin{eqnarray}
& & (1-p) \tilde{\rho}(r,v,t) \int dw \, (w-v) \tilde{\rho}(r,w,t) \nonumber \\
&\approx & (1-p) \tilde{\rho}_{(0)}(r,v,t) \rho (V - v) 
  - \int dw \, L(v,w;r,t) \tilde{\rho}_{(1)}(r,w,t) 
\end{eqnarray}
where we have introduced a linear operator $\underline{L}$ with the components
\begin{equation}
 L(v,w;r,t) := %- (1-p) \{ \rho(r,t) [V(r,t) - v] \delta(w - v) \nonumber \\
% &+& \tilde{\rho}_{(0)}(r,v,t) (w-v) \} \nonumber \\
 (1-p) \rho(r,t) \{ [v - V(r,t)]\delta(v - w) + P_{(0)}(v;r,t) (v - w) \} \, .
\end{equation}
Here, $\delta(v-w)$ denotes Dirac's delta function. The linear operator
$\underline{L}$ possesses an infinite number of eigenvalues
$1/\tau_\mu$ (cf. \cite{Lib,Uhl,Wang,Grad,Grad2}). 
The relevant eigenvalue is the smallest one 
since it characterizes temporal changes that take place on the time scale we
are interested in. It is of the order of the average {\em interaction rate}
per vehicle \cite{Rieck,Jaeckle,Huang}
\begin{eqnarray}
 \frac{1}{\tau_0} &:=& \frac{1-p}{\rho(r,t)} \int dv 
 \!\!\int\limits_{w<v}\!\! dw \, |w - v|
 \tilde{\rho}(r,w,t)\tilde{\rho}(r,v,t) \nonumber \\
 &\approx& (1-p) \rho(r,t) \int dv \!\!
 \int\limits_{w<v}\!\! dw \, |w - v| P_{(0)}(w;r,t) 
 P_{(0)}(v;r,t)  \nonumber \\ 
 &=& (1-p) \rho(r,t) \sqrt{\frac{\Theta}{\pi}} \, .
\end{eqnarray}
The other eigenvalues are somewhat larger \cite{Lib,Uhl,Wang,Grad,Grad2} 
(i.e. $\tau_\mu < \tau_0$
for $\mu \ne 0$) and they describe fast fluctuations which can be 
{\em adiabatically eliminated} \cite{Hak}. 
As a consequence, we can make the so-called
{\em `relaxation time approximation'} \cite{Anm2}
\begin{equation}
 \int dw \, L(v,w;r,t) \tilde{\rho}_{(1)}(r,w,t) \approx 
 \frac{\tilde{\rho}_{(1)} (r,v,t)}{\tau_0} \, .
\end{equation}\end{mathletters}
Now, we calculate
\begin{eqnarray}
& & \frac{\tilde{V}_0(v) - v}{\tau} \frac{\partial \tilde{\rho}_{(0)}}{\partial v}
 + \frac{\tilde{\rho}_{(0)}}{\tau} \left( \frac{\partial \tilde{V}_0(v)}
 {\partial v} - 1 \right)
 - (1-p) \tilde{\rho}_{(0)} \rho (V - v) \nonumber \\
&=& \frac{1}{\tau} \left( V_0 + \frac{{\cal C}}{\Theta} \delta v - v \right)
 \left(-\frac{\tilde{\rho}_{(0)}}{\Theta} \delta v \right)
 + \frac{\tilde{\rho}_{(0)}}{\tau} \left( \frac{{\cal C}}{\Theta} - 1 \right)
 + (1-p) \tilde{\rho}_{(0)} \rho \, \delta v \nonumber \\
&=& \frac{\tilde{\rho}_{(0)}}{\tau\Theta} \left[
 ({\cal C} - \Theta) \left( 1 - \frac{(\delta v)^2}{\Theta} \right)
 - (V_e - V) \delta v \right] \, .
\label{num5}
\end{eqnarray}
Inserting (\ref{num1}), (\ref{num2}), and (\ref{num3}) to (\ref{num5})
into the reduced Paveri-Fontana equation (\ref{Fontana}) we finally obtain
\begin{eqnarray}
 \tilde{\rho}_{(1)}(r,v,t) &\approx & - \tau_0
 \left\{ \frac{\tilde{\rho}_{(0)}}{\rho} \left(
 \delta v \frac{\partial \rho}{\partial r} - \rho
 \frac{\partial V}{\partial r} \right) \right .\nonumber \\
 & & \qquad + \frac{\tilde{\rho}_{(0)}}{\Theta}\delta v \left(
 \delta v \frac{\partial V}{\partial r} - \frac{\Theta}{\rho}
 \frac{\partial \rho}{\partial r}  - \frac{\partial \Theta}{\partial r}
 + \frac{1}{\tau} ( V_e - V ) \right) \nonumber \\
 & & \qquad + \frac{\tilde{\rho}_{(0)}}{2\Theta} \left( \frac{(\delta v)^2}{\Theta}
 - 1 \right) \left( \delta v \frac{\partial \Theta}{\partial r}
 - 2 \Theta \frac{\partial V}{\partial r}
 + \frac{2}{\tau} ( {\cal C} - \Theta ) \right) \nonumber \\
 & & \qquad - \frac{\tilde{\rho}_{(0)}}{\Theta} \left. \left[
 \frac{{\cal C} - \Theta}{\tau} \left( \frac{(\delta v)^2}{\Theta} - 1 \right)
 + \frac{V_e - V}{\tau} \delta v \right] \right\} \nonumber \\
&=& - \tilde{\rho}_{(0)}  \tau_0 \left( \frac{(\delta v)^3}{2\Theta^2} 
 - \frac{3 \, \delta v}{2 \Theta} \right) \frac{\partial \Theta}{\partial r} 
\, .
\end{eqnarray}
Obviously, the correction term $\tilde{\rho}_{(1)}(r,v,t)$ is a consequence
of the finite {\em interaction free time} $\tau_0$ which
causes a delayed adjustment of $\tilde{\rho}(r,v,t)$
to the local equilibrium $\tilde{\rho}_{(0)}(r,v,t)$. 
%and gives rise to {\em `transport terms'}
%${\cal J}$ which describe a flux density of velocity variance.
%However, this is only the case in non-equilibrium situations, namely
%if $\partial \Theta/\partial r \ne 0$.
%\par
However, in order to take into account the effects of finite {\em reaction 
time} and {\em braking time} we must add a time period $\tau' > 0$ to the
interaction free time $\tau_0$. Hence, $\tau_0$ must be replaced by the
{\em adaptation time}
\begin{equation}
 \tau_* = \tau_0 + \tau' \, .
\end{equation}

\section{Navier-Stokes-like traffic equations}

With the corrected phase-space density
\begin{eqnarray}
 \tilde{\rho}(r,v,t) &\approx & \tilde{\rho}_{(0)}(r,v,t)
 + \tilde{\rho}_{(1)}(r,v,t) \nonumber \\
 &\approx & \tilde{\rho}_{(0)}(r,v,t)
 \left[ 1 - \tau_* \left( \frac{(\delta v)^3}{2\Theta^2} 
 - \frac{3 \, \delta v}{2\Theta} \right) \frac{\partial \Theta}{\partial r}
 \right]
\end{eqnarray}
we can calculate corrected relations for the collective (`macroscopic')
quantities 
\begin{equation}
 F(r,t) \equiv \langle f(v) \rangle := \int dv \, f(v)
 \frac{\tilde{\rho}(r,v,t)}{\rho(r,t)} \approx F_{(0)}(r,t) + F_{(1)}(r,t) 
\end{equation}
where
\begin{equation}
 F_{(i)}(r,t) \equiv \langle f(v) \rangle_{(i)} := \int dv \, f(v) 
 \frac{\tilde{\rho}_{(i)}(r,v,t)}{\rho(r,t)} \, .
\end{equation}
We find
\begin{eqnarray}
 \rho(r,t) \approx  \rho_{(0)}(r,t) \, , &\qquad &
 V(r,t) \approx V_{(0)}(r,t) \, , \nonumber \\
 \Theta(r,t) \approx \Theta_{(0)}(r,t) \, , &\qquad &
 {\cal P}(r,t) \approx {\cal P}_{(0)}(r,t) \, , 
\label{gleich}
\end{eqnarray}
and
\begin{equation}
  {\cal C}(r,t) \approx {\cal C}_{(0)}(r,t) \, .
\end{equation}
However, for the flux density of velocity variance we get
\begin{eqnarray}
 {\cal J}(r,t) &\approx & {\cal J}_{(1)}(\rho,V,\Theta)
 \equiv \rho \Gamma_{(1)}(\rho,V,\Theta)
 = - \kappa \frac{\partial \Theta}{\partial r} \, ,
\label{JoT}
\end{eqnarray}
where
\begin{equation}
 \kappa := 3 \rho \tau_* \Theta 
\end{equation}
is called a {\em `kinetic coefficient'}. Therefore, the
macroscopic traffic equations (\ref{dens}), (\ref{veloc}), and (\ref{var})
assume form
\begin{equation}
 \frac{\partial \rho}{\partial t} + \frac{\partial (\rho V)}{\partial r}
 = 0 \, ,
\label{kontin}
\end{equation}
\begin{equation}
 \frac{\partial V}{\partial t} + V \frac{\partial V}{\partial r}
 = - \frac{1}{\rho} \frac{\partial {\cal P}}{\partial r}
 + \frac{1}{\tau} [ V_e(\rho,V,\Theta) - V ] \, ,
\label{geschw}
\end{equation}
and
\begin{eqnarray}
 \frac{\partial \Theta}{\partial t} + V\frac{\partial \Theta}{\partial r}
&=& - \frac{2{\cal P}}{\rho} \frac{\partial V}{\partial r} + \frac{1}{\rho}
 \frac{\partial}{\partial r} \left( \kappa \frac{\partial \Theta}{\partial r}
 \right) \nonumber \\
&+& \frac{2}{\tau} ( {\cal C} - \Theta ) 
 + (1-p) \kappa \frac{\partial \Theta}{\partial r} \, .
\label{varianz}
\end{eqnarray}
Additionally, the corrected {\em covariance equation} becomes
\begin{eqnarray}
 \frac{\partial {\cal C}}{\partial t} + V \frac{\partial {\cal C}}{\partial r}
&=& - {\cal C} \frac{\partial V}{\partial r} - \frac{{\cal P}}{\rho} 
 \frac{\partial V_0}{\partial r} + \frac{1}{\rho} \frac{\partial}{\partial r} 
   \left( \zeta \frac{\partial \Theta}{\partial r} \right) \nonumber \\
&+& \frac{1}{\tau}[{\cal C}_e(\rho,V,\Theta,{\cal C})
  - {\cal C}] + \frac{(1-p)}{2} \zeta \frac{\partial \Theta}{\partial r} 
\label{kovarianz}
\end{eqnarray}
with the kinetic coefficient
\begin{equation}
  \zeta := \kappa \frac{{\cal C}}{\Theta} = 3 \rho \tau_* {\cal C}
\end{equation}
and the {\em equilibrium covariance}
\begin{equation}
 {\cal C}_e(\rho,V,\Theta, {\cal C}) := \Theta_0 - 2 \tau
 (1-p) \rho {\cal C} \sqrt{\frac{\Theta}{\pi}} \, .
\label{eqcov}
\end{equation}
(For a detailed derivation of (\ref{kovarianz}) to (\ref{eqcov}) 
cf. Ref. \cite{hifi}.)
\par
Equations (\ref{kontin}), (\ref{geschw}), and (\ref{varianz})
are the {\em Navier-Stokes-like traffic 
equations} \cite{Rem2}. 
Compared with the Navier-Stokes equations for ordinary fluids they possess
the additional terms $(V_e - V)/\tau$ and $2(\Theta_e - \Theta)/\tau$
with $\Theta_e = {\cal C} + (\tau/2) (1-p) \kappa \partial \Theta /\partial r$
which are due to acceleration and interaction processes. 
Because of the spatial one-dimensionality of the considered traffic equations,
the velocity equation (\ref{geschw}) does not include a {\em shear viscosity
term} $(1/\rho)\partial/\partial r(\nu_0 \partial V/\partial r)$. 
The variance equation (\ref{varianz})
is related to the {\em equation of heat conduction}. However, $\Theta$
does not have the interpretation of `heat' but only of velocity variance,
here. Finally, the Navier-Stokes-like traffic equations are complemented
by the additional covariance equation (\ref{kovarianz})
arising from the tendency of drivers to get ahead
with a certain desired velocity $v_0$. 
\par
We recognize that the first-order macroscopic traffic equations
(\ref{kontin}), (\ref{geschw}), (\ref{varianz}), and 
(\ref{kovarianz}) build a closed
system of equations. Moreover, according to (\ref{gleich}), the
relations for the spatial density, average velocity, velocity variance, 
and traffic pressure did not change. 
In this sense, the chosen Chapman-Enskog
method for closing the hierarchy of macroscopic equations 
is consistent with its assumption, according to which only the expressions 
for the flux density of velocity variance ${\cal J} \equiv \rho\Gamma$ 
and the covariance ${\cal C}$ were to be improved by the non-equilibrium 
correction $\tilde{\rho}_{(1)}(r,v,t)$. However, note that another relation for
$\tilde{V}_0(v)$ than (\ref{missing}) would have led to modifications of 
$\rho$, $V$, and/or $\Theta$.
\par
We also recognize that the finite adaptation time $\tau_*$ for approaching 
the equilibrium distribution (\ref{local}) causes a finite {\em skewness}
\begin{equation}
 \gamma := \frac{\Gamma}{\Theta^{3/2}}
 = \frac{{\cal J}}{\rho \Theta^{3/2}} = - \frac{\kappa}{\rho
   \Theta^{3/2}} \frac{\partial \Theta}{\partial r} = - \frac{3 \tau_*}
 {\sqrt{\Theta}} \frac{\partial \Theta}{\partial r}
\end{equation}
of the non-equilibrium velocity distribution
\begin{equation}
 P(v;r,t) \approx \frac{\tilde{\rho}_{(0)}(r,v,t) + \tilde{\rho}_{(1)}(r,v,t)}
 {\rho(r,t)} \, .
\end{equation}
This leads to the so-called {\em transport terms}
\begin{equation}
 - \kappa \frac{\partial \Theta}{\partial r}
 \qquad \mbox{and} \qquad 
 - \zeta \frac{\partial \Theta}{\partial r} \, .
\end{equation}
The effect of these terms in equations (\ref{varianz}) and (\ref{kovarianz})
is to smooth out sudden changes of variance and covariance via 
second spatial derivatives of $\Theta(r,t)$, namely
\begin{equation}
 \frac{\partial}{\partial r} \left( \kappa \frac{\partial \Theta}{\partial r}
 \right) \qquad \mbox{and} \qquad \frac{\partial}{\partial r} \left( 
 \zeta \frac{\partial \Theta}{\partial r} \right) \, .
\end{equation}
%\begin{equation}
% - \frac{1}{\rho} \frac{\partial {\cal J}_e}{\partial r}
%%= \frac{1}{\rho} \frac{\partial}{\partial r}
%% \left( \kappa \frac{\partial \Theta}{\partial r} \right) \, .
% = \frac{\kappa}{\rho} \frac{\partial^2 \Theta}{\partial r^2}
% + \frac{1}{\rho}\frac{\partial \Theta}{\partial r} \frac{\partial
% \kappa}{\partial r} \, .  
%\end{equation}
%A similar thing holds for the covariance.

\section{Corrections of the model}

\subsection{Driver behavior and bulk viscosity}

We remember that the term $-(1/\rho)\partial {\cal P}/\partial r$ describes an
anticipation effect. It reflects that drivers accelerate when the 
`traffic pressure' ${\cal P} = \rho \Theta$ lessens, i.e. when the density
$\rho$ or the variance $\Theta$ decreases. However, drivers additionally react
to a spatial change of average velocity. 
%which also motivates them to accelerate. 
This effect can be modelled by the modified pressure relation
\begin{equation}
{\cal P}(\rho,V,\Theta) := \rho \Theta - \eta \frac{\partial V}{\partial r}
\label{pre}
\end{equation}
which gives velocity equation (\ref{geschw}) a similar form like variance
equation (\ref{varianz}) and covariance equation (\ref{kovarianz}).
\par
In order to present reasons for relation (\ref{pre}) let us assume that drivers
switch between two {\em driving modes} $m \in \{ 1, 2\}$ depending on the
traffic situation. Let $m=1$ characterize a {\em brisk}, $m=2$ describe a
{\em careful} driving mode. Then, we can split the density $\rho(r,t)$ into
{\em partial densities} $\rho_m(r,t)$ that 
delineate drivers who are in state $m$:
\begin{equation}
 \rho_{1}(r,t) + \rho_{2}(r,t) = \rho(r,t) \, .
\end{equation}
Both densities are governed by a continuity equation, but this time we have
transitions between the two driving modes with a rate
$R(\rho_{1},V)$ so that
\begin{mathletters}\label{parT}
\begin{eqnarray}
 \frac{\partial \rho_{1}}{\partial t} &=& - \frac{\partial}{\partial r}
 (\rho_{1} V) - R(\rho_{1},V) \, , \label{eins} \\
 \frac{\partial \rho_{2}}{\partial t} &=& - \frac{\partial}{\partial r}
 (\rho_{2} V) + R(\rho - \rho_{2},V) \, .
\end{eqnarray}\end{mathletters}
Adding both equations we see that the original continuity equation 
(\ref{kontin}) is still valid. Now, defining the substantial time derivative 
\begin{equation}
 \frac{D}{Dt} :=\frac{\partial}{\partial t} + V\frac{\partial}{\partial r}
\end{equation}
we can rewrite (\ref{eins}) and obtain
\begin{equation}
 \frac{D\rho_{1}}{Dt} = - \rho_{1} \frac{\partial V}{\partial r}
 - R(\rho_{1},V) \, .
\end{equation}
$D/Dt$ describes temporal changes in a coordinate system
that moves with velocity $V$. Assuming that $\rho_{1}$ relaxes rapidly 
we can apply the {\em adiabatic approximation} \cite{Hak}
\begin{equation}
 \frac{D\rho_{1}}{Dt} \approx 0 \label{adel}
\end{equation}
which is valid on the slow time-scale of the macroscopic changes of traffic
flow. This leads to 
\begin{equation}
 R(\rho_{1},V) \approx - \rho_{1} \frac{\partial V}{\partial r} \, .
\label{inc}
\end{equation}
Relation (\ref{adel}) implies that the density $\rho_{1}$ of 
briskly behaving drivers is approximately constant in the moving coordinate
system whereas the density $\rho_{2} = \rho - \rho_{1}$ of carefully behaving
drivers varies with the traffic situation:
\begin{equation}
 \frac{D\rho_{2}}{Dt} \approx - \rho \frac{\partial V}{\partial r} \, . 
\end{equation}
$\rho_{2}$ increases when
the average velocity spatially decreases ($\partial V/\partial r < 0$)
since this may indicate a critical traffic situation.
\par
According to relations (\ref{parT}), 
(\ref{inc}) incessant transitions between the two driving
modes take place as long as traffic flow is spatially non-homogeneous
(i.e. $\partial V/\partial r \ne 0$). This leads to corrections of the
pressure relation. Expanding ${\cal P}$ with
respect to the variable $R$ which characterizes the disequilibrium between
the two driving modes we find \cite{Keiz}
\begin{equation}
 {\cal P}(\rho,\Theta,R) 
 = {\cal P}(\rho,\Theta,0) - \left. \frac{\partial {\cal P}}
 {\partial R} \right|_{R=0} \rho_{1} \frac{\partial V}{\partial r} + \dots \, .
\end{equation}
With the equilibrium relation ${\cal P}(\rho,\Theta,0) = \rho \Theta$
and
\begin{equation}
 \eta := \rho_{1} \left. \frac{\partial {\cal P}}{\partial R}\right|_{R=0}
\end{equation}
we finally obtain the desired result
\begin{equation}
 {\cal P}(\rho,\Theta,R) \equiv {\cal P}(\rho,V,\Theta)
 = \rho \Theta - \eta \frac{\partial V}{\partial r} \, .
\end{equation}
A more detailed discussion can be found in Ref. \cite{Keiz}.

\subsection{Modifications due to finite space requirements}

We will now introduce some corrections that are due to the fact that
vehicles are no point-like objects but need, on average, a space of
\begin{equation}
 s(V) = l + VT
\end{equation}
each. Here, $l\ge l_0$ is about the {\em average vehicle length} whereas $VT$ 
corresponds to the {\em safe distance} each driver should keep to the next
vehicle ahead. $T$ is about the {\em reaction time}. Consequently, if
$\Delta N(r,t) := \rho(r,t) \, \Delta r$ means the number of vehicles that are
at a place between $r - \Delta r/2$ and $r + \Delta r/2$, the {\em effective
density} is
\begin{equation}
 \varrho(r,t) = \frac{\Delta N(r,t)}{\Delta r - \Delta N(r,t) s[V(r,t)]}
 = \frac{\rho(r,t)}{1 - \rho(r,t)s[V(r,t)]} \, .
\end{equation}
Since $\Delta N(r,t) s(V)$ is the space which is occupied by $\Delta N(r,t)$
vehicles, the effective density is the number $\Delta N(r,t)$ of vehicles
per {\em effective free space} $\Delta r - \Delta N(r,t) s(V)$. 
\par
The reduction
of available space by the vehicles leads to an increase of their interaction
rate. Therefore, we have
\begin{eqnarray}
 \left( \frac{\partial \hat{\rho}}{\partial t} \right)_{\rm tr}
 &:=& (1-p) \int\limits_v^\infty dw \int dw_0 \, |v - w| \hat{\varrho}(r,v,w_0,t)
 \hat{\rho}(r,w,v_0,t) \nonumber \\
&-& (1-p) \int\limits_0^v dw \int dw_0 \, |w - v| \hat{\varrho}(r,w,w_0,t)
 \hat{\rho}(r,v,v_0,t)
\end{eqnarray}
with
\begin{equation}
 \hat{\varrho}(r,v,v_0,t) := \frac{\hat{\rho}(r,v,v_0,t)}
 {1 - \rho(r,t) s[V(r,t)]} \, .
\end{equation}
Consequently, we obtain the corrected relation
\begin{equation}
 \frac{1}{\tau_0} := (1-p) \varrho \sqrt{\frac{\Theta}{\pi}} \, .
\end{equation}
In addition, we must replace ${\cal P}$ and ${\cal J}$ by
\begin{equation}
 {\cal P}' := \frac{{\cal P}}{1 - \rho s(V)}
\qquad \mbox{and} \qquad {\cal J}' := \frac{{\cal J}}{1 - \rho s(V)} 
\end{equation}
respectively \cite{Klim}. For the kinetic coefficients
$\eta$, $\kappa$, and $\zeta$ we obtain the corrected relations
%\begin{equation}
% {\cal P}' = \varrho \Theta - \zeta' \frac{\partial V}{\partial r}
%\qquad \mbox{and} \qquad
% {\cal J}' = - \kappa' \frac{\partial \Theta}{\partial r}
%\end{equation}
%with
\begin{displaymath}
 \eta' := \frac{\eta}{1 - \rho s(V)} \, , \qquad
 \kappa' := \frac{\kappa}{1 - \rho s(V)} = 3 \varrho \tau_* \Theta \, ,
\end{displaymath}
\begin{equation}
 \mbox{and} \qquad 
 \zeta' := \frac{\zeta}{1 - \rho s(V)} = 3 \varrho \tau_* {\cal C} \, .
\end{equation}
The corrected formula 
\begin{equation}
 \varrho \Theta = \frac{\rho \Theta}{1 - \rho s(V)}
\label{Waals}
\end{equation}
for the equilibrium pressure corresponds to the pressure relation
of van der Waals for a `real gas'.
According to (\ref{Waals}), the traffic
pressure diverges for $\rho \rightarrow \rho_{\rm max} := 1/l$ which causes
a deceleration of vehicles.
\par
The corrected kinetic coefficients $\eta'(\rho,V,\Theta)$,
$\kappa'(\rho,V,\Theta)$, and
$\zeta'(\rho,V,\Theta,{\cal C})$ also diverge for 
$\rho \rightarrow \rho_{\rm max}$ \cite{Klim}. 
We find for example
\begin{equation}
 \kappa' \stackrel{\rho \approx \rho_{\rm max}}{\longrightarrow}
 3 \varrho \tau' \Theta = \frac{3 \rho \tau' \Theta}{1 - \rho s(V)}
\label{modi2}
\end{equation}
so that the divergence of $\kappa'$ is a consequence of the finite
reaction- and braking-time $\tau'$. 
This divergence causes a homogenization
of traffic flow since the second spatial derivatives 
$\partial/\partial r( \eta \partial V/\partial r)$, 
$\partial/\partial r( \kappa \partial \Theta/\partial r)$, 
and $\partial/\partial r( \zeta \partial \Theta/\partial r)$ produce a
spatial smoothing of average velocity $V$, variance $\Theta$,
and covariance ${\cal C}$ respectively.
%This can be imagined
%in the following way: At high densities vehicles are queuing.
%This implies that most vehicles move with the same
%velocity over rather long distances leading to a spatial homogenization of
%traffic conditions. 
%However, in the Navier-Stokes-like equations this homogenization is
%described by the second spatial derivatives $\zeta \partial^2 V/\partial
%r^2$ and $\kappa \partial^2 \Theta/\partial r^2$ since these produce a
%spatial smoothing of average velocity $V$ and variance $\Theta$ respectively.
%The magnitude of $\zeta$ and $\kappa$ is a measure for the strength of this
%homogenization effect.
\par
It is the divergence of `traffic pressure' and kinetic coefficients
for $\rho \rightarrow \rho_{\rm max}$
that prevents the spatial density $\rho$ from exceeding the maximum density
$\rho_{\rm max}$ \cite{Hel1}.

\section{Summary and outlook}

This paper started with a discussion of the most widespread macroscopic traffic
models. Each of them is suitable for the description of certain
traffic situations on freeways but fails for others. Therefore, an improved
fluid-dynamic model was derived from the gas-kinetic traffic equation
of Paveri-Fontana \cite{Pav} which is very well justified and does not show the
peculiar properties of Prigogine's Boltzmann-like approach \cite{Prig}.
\par
For the derivation of the improved traffic model,
moment equations for collective ('macroscopic') quantities like the
spatial density, average velocity, and velocity variance
had to be calculated. The system of macroscopic equations turned
out to be non-closed so that a suitable approximation was necessary. Here,
the well proved Chapman-Enskog method was applied. In zeroth-order
approximation the velocity distribution is assumed to be in 
`local equilibrium'. According to empirical data, the latter is characterized by
a Gaussian velocity distribution. Depending on the respective kind of zeroth-order
approximation one arrives at the Lighthill-Whitham model \cite{LW},
the model of Phillips \cite{Phil,Phil2}, or the Euler-like traffic equations.
\par
For the derivation of a first-order approximation, %we needed stationary and
%spatially homogeneous solution of the reduced Paveri-Fontana's equation. 
%In accordance with empirical traffic data a Gaussian velocity distribution
%was taken. Afterwards, an approximate time-dependent
%solution for the velocity distribution $P(v;r,t)$ was obtained from
the reduced Paveri-Fontana equation was linearized around the local
equilibrium solution and solved by application of the Euler-like traffic
equations. The resulting correction term for the non-equilibrium
velocity distribution allowed the calculation of additional
transport terms which describe a flux density of velocity variance and covariance in 
spatially non-homogeneous situations. They are related
with a finite skewness of the velocity distribution. The shear-viscosity
term vanishes because of the one-dimensionality of the considered traffic
equations. Nevertheless, a bulk-viscosity term results from transitions 
between two different driving modes: a brisk and a careful one.
\par
The resulting Navier-Stokes-like traffic equations were 
finally corrected in order to
take into account the finite space requirements of vehicles.
They overcome the deficiencies of the former macroscopic traffic models
so that the criticism by Daganzo \cite{Daganzo} and others could be invalidated:
\begin{itemize}
\item[1.] The anticipation term which, in other models, is responsible for the
prediction of negative velocities vanishes in problematic situations like the 
one described at the end of Sec. II since the variance becomes 
zero, then.
\item[2.] The density $\rho(r,t)$ does not exceed the 
maximum admissible density
$\rho_{\rm bb}$ (= bumper-to-bumper density) \cite{Hel1} since the diverging
viscosity term causes a homogenization of traffic flow and the diverging
traffic pressure suppresses an unrealistic growth of velocity which stops 
a further increase of traffic density.
\item[3.] The model takes into account different driving styles by a
distribution of desired velocities $v_0$ which are directly associated with the
individual drivers. An extension of the Navier-Stokes-like traffic equations to
different vehicle types (cars and trucks) is possible \cite{Hel4}.
\item[4.] The interaction between drivers is modelled 
anisotropically since the slower vehicle is assumed not to be affected by a
faster vehicle behind it or overtaking it.
\item[5.] According to the Navier-Stokes-like equations, disturbances may
propagate with a velocity $c > V$ since a certain proportion of vehicles
moves faster than the average velocity $V$ due to the finite velocity variance
$\Theta$. Therefore, in contrast to what was claimed by
Daganzo \cite{Daganzo}, it is admissible that macroscopic traffic models
``exhibit one characteristic speed greater than the macroscopic fluid
velocity'' \cite{Daganzo,Anm3}. %However, it should certainly hold
%$c \le V + 3 \sqrt{\Theta}$.
\end{itemize}
\par
Present investigations focus on the {\em computer simulation} of the
Navier-Stokes-like traffic equations. This work has already been successfully
started for a circular road \cite{Hel1,Anm4} and is now extended to 
complex freeway networks.
\par
Moreover, the gas-kinetic and Navier-Stokes-like traffic models can be 
generalized to models for
multi-lane traffic where overtaking and lane-changing is explicitly taken into
account \cite{Hel4}. By this, formulas for the relations
$\tau(\rho,V,\Theta)$ and $p(\rho,V,\Theta)$ can be derived \cite{Bemerk2}.

\section*{Acknowledgements}

The author wants to thank M. Hilliges, R. K\"uhne, and P. Konh\"auser
for valuable and inspiring discussions.

%\end{multicols}

\begin{references}
\bibitem{Rem2} Often the terms `Euler equation' and `Navier-Stokes equation'
denote the respective velocity equation only. Since it is easier to name
the whole set of connected equations (containing the respective
density equation, velocity equation, and variance equation)
by one single term, the notation of Ref. \cite{Lib} has been applied, here.
\bibitem{Notiz} Phillips \cite{Phil} fits his traffic data not to a Gaussian
distribution but to a more complicated one which he derived from
his gas-kinetic model. However, Phillips' figures
clearly show that, in the region of stable traffic, 
the asymmetric distribution function favoured by him
does not fit the data as well as a Gaussian distribution would.
\bibitem{Anm2} To illustrate this, assume that we are confronted with  a linear
differential equation
$d\vec{x}/dt = - \underline{L} \vec{x} $
with a linear operator (or matrix) $\underline{L}$. This equation has the
general solution $\vec{x}(t) = \sum_\mu a_\mu \vec{x}_\mu 
\mbox{e}^{-t/\tau_\mu}$
where the eigenvalues $1/\tau_\mu$ are the solutions of the characteristic
equation $ \mbox{det} ( \underline{L} - 1/\tau_\mu ) = 0 $
and the eigenvectors $\vec{x}_\mu$ satisfy the equation
$ \underline{L} \vec{x}_\mu = \vec{x}_\mu/\tau_\mu$.
If $1/\tau_\mu \gg 1/\tau_0$ for $\mu \ne 0$ we have
$\vec{x}(t) \approx a_0 \vec{x}_0 \mbox{e}^{-t/\tau_0}$ 
after a very short time $ t > 3 \max \{ \tau_\mu : \mu \ne 0 \} $
(adiabatic approximation). This implies
$ - \underline{L} \vec{x}(t) = d\vec{x}(t)/dt \approx 
 - \vec{x}(t)/\tau_0$.
\bibitem{Anm3} Here, the term ``characteristic speeds'' refers to
the slopes of the characteristics in the $(r,t)$-plane which can be calculated
for a hyperbolic system of partial differential equations \cite{Whith}.
\bibitem{Anm4} However, note that the traffic model simulated in
  Ref. \cite{Hel1}
differs in some details from the Navier-Stokes-like model derived in this
paper.
\bibitem{Bemerk2} If one assumes $\tau(\rho,V,\Theta) \equiv \tau(\rho)$
and $p(\rho,V,\Theta) \equiv p(\rho)$ like Prigogine \cite{Prig}, Phillips
\cite{Phil}, and Paveri-Fontana \cite{Pav} did, the unknown functional
relation for the product $\tau(\rho)[1 - p(\rho)]$ can be determined
from the empirical velocity-density relation $V_e(\rho_e)$ by inserting
(\ref{Zwei}) into (\ref{Eins}).
\end{references}
\end{document}